\documentclass
[superscriptaddress,12pt,a4paper,preprint,showpacs,showkeys,nobibnotes,nofootinbib]{revtex4}%
\usepackage{amsfonts}
\usepackage{amsmath}
\usepackage{amssymb}
\usepackage{graphics}
\usepackage{graphicx}%
\providecommand{\U}[1]{\protect\rule{.1in}{.1in}}
\begin{document}
\author{Mario Castagnino}
\affiliation{CONICET - IAFE - IFIR - Universidad de Buenos Aires, Argentina}
\author{Sebastian Fortin}
\affiliation{CONICET - IAFE - Universidad de Buenos Aires, Argentina}
\email{sfortin@gmx.net}

\begin{abstract}
There are many formalisms to describe quantum decoherence. However, many of
them give a non general and \ ad hoc definition of \textquotedblleft pointer
basis\textquotedblright\ or \textquotedblleft moving preferred
basis\textquotedblright, and this fact is a problem for the decoherence
program. In this paper we will consider quantum systems under a general
theoretical framework for decoherence and present a very general definition of
the moving preferred basis. In addition, this definition is implemented in a
well known model and the time of decoherence and the relaxation time are
defined and compared with those of this model.

\end{abstract}
\title{Defining the moving preferred basis}

\maketitle
\tableofcontents

\section{Introduction}

From the appearance of the quantum mechanics many attempts have been made to
recover the laws of the classic mechanics through some classic limit. The more
common scheme of this type includes the \textit{quantum decoherence}%
\footnote{We will call \textit{decoherence} to the vanishing of the
off-diagonal terms in a properly specified basis. We will call
\textit{relaxation} to the decoherence in a final equilibrium basis, i.e.
typical equilibrium}. This process is in charge to erase the terms of
interference of the density matrix, that are classically inadmissible, since
they prevent the use of a classical (boolean) logic. In addition, decoherence
leads to the rule that selects the candidates for classic states.

As it is pointed out in the brief historical summary of paper
\cite{CQG-CFLL-08}, three periods can be schematically identified in the
development of the general program of decoherence \cite{Omnes}. A first
period, when the arrival to the equilibrium of irreversible systems was
studied. During this period, authors as van Kampen, van Hove, Daneri, et
al\textit{.} developed a formalism for explaining the decoherence phenomenon
that was not successful at the time but it established the bases of this
study. The main problem of this period was that too long decoherence times
were found, if compared with the experimental ones. In a second period the
decoherence in open systems was studied, the main characters of this period
were Zeh and Zurek. In their works, the decoherence is an interaction process
between an open quantum system and its environment. This process, called
\textit{Environment-Induced Decoherence} (EID), determines, case by case,
which is the privileged basis, usually called \textit{moving preferred basis}
where decoherence takes place in a \textit{decoherence time} $t_{D}$ and it
defines the observables that acquire classic characteristics and they could be
interpreted in some particular cases as properties that obey a Boolean logic.
This is the orthodox position in the subject \cite{Bub}. The decoherence times
in this period were much smaller, solving the problem of the first period.
Recently, in a third period it becomes evident that dissipation was not a
necessary condition for decoherence \cite{Max} and the study of the arrival to
equilibrium of closed systems was also considered. We will not discuss closed
systems in this paper but for the sake of completeness we will make only some
comments. Closed system will be discussed at large elsewhere.

In this work we focus the attention on EID, which is a well known theory, with
well established experimental verifications, which makes unnecessary any
further explanation. On the contrary other formalisms are not so well
established, but they must be taken into account for the sake of completeness
(\cite{GP}, \cite{connel}, \cite{CP}, \cite{SID}, \cite{SID'}, \cite{DT},
\cite{MHI}, \cite{PLA}, \cite{Studies}, \cite{JPA2}).

In this paper, we will introduce a tentative definition (for EID and other
formalisms) of moving preferred basis where the state decoheres in a very
short time $t_{D}$, So the main problem of the first period is solved in a
convenient and general way. Our main aim is to present a new conceptual
perspective that will clarify some points that still remain rather obscure in
the literature on the subject, e. g. the definition of the moving preferred basis.

\subsection{The General Theoretical Framework for Decoherence}

In previous works we have resumed the common characteristics of the different
approaches of decoherence, which suggest the existence of a general framework
for decoherence within which these approaches can all be framed (see
\cite{CQG-CFLL-08}, \cite{Studies} and \cite{MPLA}). According to this general
framework, that was developed in \cite{Studies} and will be completed in
future papers, decoherence is just a particular case of the general problem of
irreversibility in quantum mechanics. Since the quantum state $\rho(t)$
follows a unitary evolution, it cannot reach a final equilibrium state for
$t\rightarrow\infty$. Therefore, if the non-unitary evolution towards
equilibrium is to be accounted for, a further element has to be added to this
unitary evolution. The way to introduce this non-unitary evolution must
include the splitting of the whole space of observables $\mathcal{O}$ into the
relevant subspace $\mathcal{O}_{R}\subset\mathcal{O}$ and the irrelevant
subspace. Once the essential role played by the selection of the relevant
observables is clearly understood, the phenomenon of decoherence can be
explained in four general steps:

\begin{enumerate}
\item \textbf{First step:} The space $\mathcal{O}_{R}$ of relevant observables
is defined .

\item \textbf{Second step:} The expectation value $\langle O_{R}\rangle
_{\rho(t)}$, for any $O_{R}\in\mathcal{O}_{R}$, is obtained. This step can be
formulated in two different but equivalent ways:

\begin{itemize}
\item A coarse-grained state $\rho_{R}(t)$ is defined by
\begin{equation}
\langle O_{R}\rangle_{\rho(t)}=\langle O_{R}\rangle_{\rho_{R}(t)} \label{0}%
\end{equation}
for any $O_{R}\in\mathcal{O}$, and its non-unitary evolution (governed by a
master equation) is computed (this step is typical in EID).

\item $\langle O_{R}\rangle_{\rho(t)}$ is computed and studied as the
expectation value of $O_{R}$ in the state $\rho(t)$. This is the generic case
for other formalisms.
\end{itemize}

\item \textbf{Third step:} It is proved that $\langle O_{R}\rangle_{\rho
(t)}=\langle O_{R}\rangle_{\rho_{R}(t)}$ reaches a final equilibrium value
$\langle O_{R}\rangle_{\rho_{\ast}}$, then
\begin{equation}
\lim_{t\rightarrow\infty}\langle O_{R}\rangle_{\rho(t)}=\langle O_{R}%
\rangle_{\rho_{\ast}},\text{ \ \ \ \ \ \ \ \ \ \ \ }\forall O_{R}%
\in\mathcal{O}_{R} \label{INT-01}%
\end{equation}

This also means that the coarse-grained state $\rho_{R}(t)$ evolves towards a
final equilibrium state:%
\begin{equation}
\lim_{t\rightarrow\infty}\langle O_{R}\rangle_{\rho_{R}(t)}=\langle
O_{R}\rangle_{\rho_{R\ast}},\ \ \ \ \ \ \ \ \ \ \ \forall O_{R}\in
\mathcal{O}_{R} \label{INT-02'}%
\end{equation}

The characteristic time for these limits is the $t_{R}$, the
\textit{relaxation time.}

\item \textbf{Fourth step: }Also a \textit{moving preferred basis}
$\{|\widetilde{j(t)\rangle}\}$ must be defined as we will see in section I.B.
This basis is the eigen basis of certain state $\rho_{P}(t)$ such that%
\begin{equation}
\lim_{t\rightarrow\infty}\langle O_{R}\rangle_{(\rho_{R}(t)-\rho_{P}%
(t))}=0,\ \ \ \ \ \ \ \ \ \ \ \forall O_{R}\in\mathcal{O}_{R}%
\end{equation}
The characteristic time for this limit is the $t_{D}$, the \textit{decoherence
time.}
\end{enumerate}

The final equilibrium state $\rho_{\ast}$ is obviously diagonal in its own
eigenbasis, which turns out to be the final preferred basis. But, from eqs.
(\ref{INT-01}) or (\ref{INT-02'}) we cannot say that $\lim_{t\rightarrow
\infty}\rho(t)=\rho_{\ast}$ or $\lim_{t\rightarrow\infty}\rho_{R}%
(t)=\rho_{R\ast}.$ Then, the mathematicians say that the unitarily evolving
quantum state $\rho(t)$ of the whole system \textit{only has a} \textit{weak
limit, }symbolized as:
\begin{equation}
W-\lim_{t\rightarrow\infty}\rho(t)=\rho_{\ast} \label{INT-03}%
\end{equation}
equivalent to eq. (\ref{INT-01}). As a consequence, the coarse-grained state
$\rho_{R}(t)$ also has a weak limit, as follows from eq.(\ref{INT-02'}):
\begin{equation}
W-\lim_{t\rightarrow\infty}\rho_{R}(t)=\rho_{R\ast} \label{INT-04}%
\end{equation}
equivalent to eq. (\ref{INT-02'}). Also%
\begin{equation}
W-\lim_{t\rightarrow\infty}(\rho_{R}(t)-\rho_{P}(t))=0
\end{equation}
These weak limits mean that, although the off-diagonal terms of $\rho(t)$
never vanish through the unitary evolution, the system decoheres \textit{from
an observational point of view}, that is, from the viewpoint given by any
relevant observable $O_{R}\in\mathcal{O}_{R}$.

From this general perspective, the phenomenon of destructive interference,
that produced the decoherence phenomenon, is relative because the off-diagonal
terms of $\rho(t)$ and $\rho_{R}(t)$ vanish only from the viewpoint of the
relevant observables\textbf{ }$O_{R}\in\mathcal{O}_{R}$, and the
superselection rule that precludes superpositions only retains the states
defined by the corresponding decoherence bases as we will see. The only
difference between EID and other formalisms for decoherence is the selection
of the relevant observables (see \cite{CQG-CFLL-08} for details):

\begin{description}
\item[.] In EID the relevant observables are those having the following form:%
\begin{equation}
O_{R}=O_{S}\otimes I_{E}\in\mathcal{O}_{R} \label{EID}%
\end{equation}
where $O_{S}$ are the observables of the system and $I_{E}$ is the identity
operator of the environment. Then eq. (\ref{0}) reads%
\[
\langle O_{R}\rangle_{\rho(t)}=\langle O_{R}\rangle_{\rho_{R}(t)}=\langle
O_{S}\rangle_{\rho_{S}(t)},\text{ where }\rho_{S}(t)=Tr_{E}\rho(t)
\]

\item[.] In the other formalisms other restriction in the set of observables
are introduced
\end{description}

\subsection{The definition of moving preferred basis}

The moving preferred basis was introduced, case by case in several papers (see
\cite{Max}) in a non systematic way. On the other hand in references
\cite{OmnesPh} and \cite{OmnesRojo} Roland Omn\`{e}s introduces a rigorous and
almost general definition of the moving preferred basis based in a reasonable
choice of the relevant observables, and other physical considerations.

In this paper we will introduce an alternative general definition to define
this basis: As it is well known the eigen values of the Hamiltonian are the
inverse of the characteristic frequencies of the unitary evolution of an
oscillatory system. Analogously, for non-unitary evolutions, the poles of the
complex extension of the Hamiltonian are the \textit{catalogue} of the
decaying modes of these non-unitary evolutions towards equilibrium (see
\cite{JPA}). This will be the main idea to implement the definition of our
moving preferred basis. i. e. we will use these poles.

We will compare and try to unify these two methods in the future. Really we
already began this approach with Omn\`{e}s in section III.

\subsection{Organization of the paper.}

In Section I we have introduced a general framework for decoherence. A general
candidate for moving decoherence basis is introduced in section
\ref{GenDefMovDecBas}, which is implemented in three toy models and the time
of decoherence and the relaxation time in these approaches are defined. In
principle these definitions can be used in EID and probably for generic
formalisms. In Section III we will present the paradigmatic EID: Omn\`{e}s (or
Lee-Friedrich) model. and show that the pole method yields the same results.
Finally in Section \ref{Conclusions} we will draw our conclusions. One
appendix completes this paper.

\section{\label{GenDefMovDecBas}Towards a general definition for the moving
preferred basis.}

\subsection{Introduction and review}

In this section we will try to introduce a very general theory for the moving
preferred basis for \textit{any relevant observable space} $\mathcal{O}_{R}.$
Then it is necessary to endow the coordinates of observables and states in the
Hamiltonian basis $\{|\omega\rangle\}$ (i.e. the functions $O(\omega
,\omega^{\prime})$ and $\rho(\omega.\omega^{\prime}))$ with extra analytical
properties in order to find the definition of a moving preferred basis in the
most, general, convincing, and simplest way. It is well known that this move
is usual in many chapters of physic e. g. in the scattering theory (see
\cite{Bohm}).

It is also well known that evolution towards equilibrium has two phases.

i.- A exponential dumping phase that can be described studying the analytical
continuation of the Hamiltonian into the complex plane of the energy (see
\cite{JPA}, \cite{Clifton}, \cite{Weder}, \cite{Sudarsham}, \cite{PRA1},
\cite{PRA2}, \cite{PRE}), a fact which is also well known in the scattering theory.

ii.- A final polynomial decaying in $t$ known as the long time of Khalfin
effect (see \cite{Khalfin}, \cite{BH}), which is very weak and difficult to
detect experimentally (see \cite{KhalfinEx}).

These two phases will play an important role in the definition of the moving
preferred basis. They can be identified by the theory of analytical
continuation of vectors, observables and states. To introduce the main
equation we will make a short abstract of papers \cite{JPA} and \cite{PRA2}.

\subsection{Analytic continuations in the bra-ket language.}

We begin reviewing the analytical continuation for pure states. Let the
Hamiltonian be $H=H_{0}+V$ where the free Hamiltonian $H_{0}$ satisfies (see
\cite{JPA}. eq. (8) or \cite{PRA2})%

\[
H_{0}|\omega\rangle=\omega|\omega\rangle,\text{ }\langle\omega|H_{0}%
=\omega\langle\omega|,\text{ \ \ }0\leq\omega<\infty
\]
and (see \cite{JPA}. eq. (9))%
\begin{equation}
I=\int_{0}^{\infty}d\omega|\omega\rangle\langle\omega|,\text{ }\langle
\omega|\omega^{\prime}\rangle=\delta(\omega-\omega^{\prime}) \label{I}%
\end{equation}
Then (see \cite{JPA}. eq. (10))
\[
H_{0}=\int_{0}^{\infty}\omega|\omega\rangle\langle\omega|d\omega
\]
and (see \cite{JPA}. eq. (11))
\begin{equation}
H=H_{0}+V=\int_{0}^{\infty}\omega|\omega\rangle\langle\omega|d\omega+\int
_{0}^{\infty}d\omega\int_{0}^{\infty}d\omega^{\prime}V_{\omega\omega^{\prime}%
}|\omega\rangle\langle\omega^{\prime}|=\int_{0}^{\infty}\omega|\omega
^{+}\rangle\langle\omega^{+}|d\omega\label{Hamil}%
\end{equation}
where the $|\omega^{+}\rangle$ are the eigenvectors of $H$, that also satisfy
eq. (\ref{I}). The eigen vectors of $H$ are given by the Lippmann-Schwinger
equations (see \cite{JPA}. eq. (12) and (13))%
\begin{align}
\langle\psi|\omega^{+}\rangle &  =\langle\psi|\omega\rangle+\langle\psi
|\frac{1}{\omega+i0-H}V|\omega\rangle\nonumber\\
\langle\omega^{+}|\varphi\rangle &  =\langle\omega|\varphi\rangle
+\langle\omega|V\frac{1}{\omega-i0-H}|\varphi\rangle\label{AN}%
\end{align}
Let us now endow the function of $\omega$ with adequate analytical properties
(see \cite{Bohm}). E.g. let us consider that the state $|\varphi\rangle$
(resp. $\langle\psi|)$ is such that it does not create poles in $\langle
\omega|\varphi\rangle$ (resp. in $\langle\psi|\omega\rangle)$ and therefore
this function is analytic in the whole complex plane. This is a simplification
that we will be forced to abandon in realistic cases as we will see. Moreover
we will consider that the function $\langle\omega^{+}|\varphi\rangle$ (resp.
$\langle\psi|\omega^{+}\rangle)$ is analytic but with just one simple pole at
$z_{0}$ $=\omega_{0}-\frac{i}{2}\gamma_{0},$ $\gamma_{0}>0$ in the lower
halfplane (resp. another pole $z_{0}^{\ast}=\omega_{0}+\frac{i}{2}\gamma
_{0},\gamma_{0}>0$ on the upper halfplane$)$ (see \cite{DT} for details
\footnote{This is a toy model with just one pole and the Khalfin effect. More
general models , with two poles, will be considered in the next subsection.
The pole corresponds to the residue that we can compute with the curve
\textit{C} and the Khalfin effect to the integral along the curve $\Gamma$ of
Figure 1.}). There can be many of such poles but , by now, we will just
consider one pole for simplicity, being the generalization straightforward.
Then we make an analytic continuation of the positive $\omega$ axis to the
curve $\Gamma$ of the lower half-plane as in Figure 1.

\begin{figure}[t]
\centerline{\scalebox{0.7}{\includegraphics{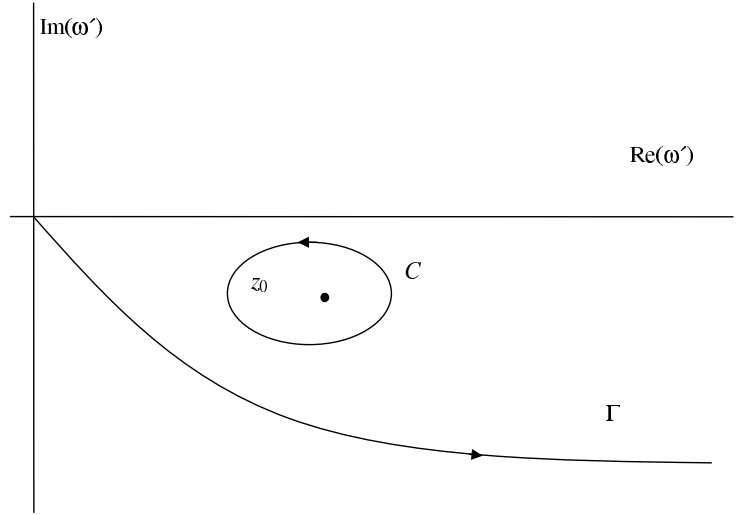}}} \vspace
*{0.cm}\caption{Complex contour $\Gamma$ on the lower complex energy plane
usedin our evaluation of integrals. The \textquotedblleft
energy\textquotedblright\ $z_{0}$ is the pole that we assume to be simple.}%
\label{fig 1}%
\end{figure}

Then (see \cite{JPA}. eq. (29)) we can define%
\begin{align}
\langle\widetilde{f_{0}}|\varphi\rangle &  \equiv cont_{\omega^{\prime
}\rightarrow z_{0}}\langle\omega^{\prime+}|\varphi\rangle\nonumber\\
\langle\psi|f_{0}\rangle &  \equiv(-2\pi i)cont_{\omega^{\prime}\rightarrow
z_{0}}(\omega^{\prime}-z_{0})\langle\psi|\omega^{+}\rangle\nonumber\\
\langle\widetilde{f_{z^{\prime}}}|\varphi\rangle &  \equiv cont_{\omega
^{\prime}\rightarrow z^{\prime}}\langle\omega^{\prime+}|\varphi\rangle
,\text{\ }z^{\prime}\in\Gamma,\forall\text{ }|\varphi\rangle\text{ }%
\langle\psi|\nonumber\\
\langle\psi|f_{z^{\prime}}\rangle &  \equiv cont_{\omega^{\prime}\rightarrow
z}\langle\psi|\omega^{+}\rangle,\text{\ }z^{\prime}\in\Gamma,\forall\text{
}|\varphi\rangle\text{ }\langle\psi| \label{L.1}%
\end{align}
and (see \cite{JPA}. eq. (31))%
\begin{align}
\langle\psi|\widetilde{f_{0}}\rangle &  \equiv cont_{\omega\rightarrow
z_{0}^{\ast}}\langle\psi|\omega^{+}\rangle\nonumber\\
\langle f_{0}|\varphi\rangle &  \equiv(2\pi i)cont_{\omega^{\prime}\rightarrow
z_{0}^{\ast}}(\omega-z_{0})\langle\omega^{+}|\varphi\rangle\nonumber\\
\langle\psi|\widetilde{f_{z^{\prime}}}\rangle &  \equiv cont_{\omega
\rightarrow z}\langle\psi|\omega^{+}\rangle,\text{ \ }z\in\Gamma,\forall\text{
}|\varphi\rangle\text{ }\langle\psi|\nonumber\\
\langle f_{z}|\varphi\rangle &  \equiv cont_{\omega\rightarrow z}\langle
\omega^{+}|\varphi\rangle,\text{ \ }z\in\Gamma,\forall\text{ }|\varphi
\rangle\text{ }\langle\psi| \label{L.2}%
\end{align}
where $cont$ means analytic continuation.

Finally it can be proved that (see \cite{JPA})
\[
H=z_{0}|f_{0}\rangle\langle\widetilde{f_{0}}|+\int_{\Gamma}z|f_{z}%
\rangle\langle\widetilde{f_{z}}|dz
\]
a simple extension of the eigen-decomposition of $H$ to the complex plane

\subsection{Analytical continuation in the observables and states language.}

We could repeat what we have said about the pure states and the Hamiltonian
with the states, observables, and the Liouvillian operator $\ L$ (see a review
in \cite{Cos}). But we prefer to follow the line of \cite{JPA} and keep the
Hamiltonian framework and discuss the analytical continuation of \ $\langle
O\rangle_{\rho(t)},$ that we will also symbolize as $(\rho(t)|O)$. In fact
from section I.A we know that this scalar is the main character so we will
study its analytical properties ad nauseam.

So let us call (see \cite{JPA}. eq. (42))
\[
|\omega)=|\omega\rangle\langle\omega|,\text{ and }|\omega,\omega^{\prime
})=|\omega\rangle\langle\omega^{\prime}|
\]
Then a generic relevant observable is $O_{R}\in\mathcal{O}_{R}$ (see
\cite{PRA2} eq. (42) or \cite{JPA}. eq. (42))%
\begin{equation}
O_{R}=|O_{R})=\int d\omega O(\omega)|\omega)+\int d\omega\int d\omega^{\prime
}O(\omega,\omega^{\prime})|\omega,\omega^{\prime}) \label{A}%
\end{equation}
and the generic states is (\cite{PRA2} eq. (45) or \cite{JPA}. eq. (45) )%
\begin{equation}
\rho_{R}=(\rho_{R}|=\int d\omega\rho(\omega)\widetilde{(\omega|}+\int
d\omega\int d\omega^{\prime}\rho(\omega,\omega^{\prime})\widetilde
{(\omega,\omega^{\prime}|} \label{B}%
\end{equation}
where $\widetilde{(\omega|},$ $\widetilde{(\omega,\omega^{\prime}|}$ are
defined those of eqs. (\ref{53'}) and (\ref{54}) in the case $V=0$ (see also
\cite{PRA2} eq. (44) or \cite{JPA}. eq. (45))%
\[
\widetilde{(\omega|}O_{R})=O(\omega),\text{ }\widetilde{(\omega,\omega
^{\prime}|}O_{R})=O(\omega,\omega^{\prime})
\]
We will keep the treatment as general as possible, i.e. $O_{R}$ would be any
observable such that $O_{R}\in\mathcal{O}_{R}$\ and $\rho_{R}$ any state
$\rho_{R}\in\mathcal{O}_{R}^{\prime}$ \footnote{Namely, even more general than
the choice of EID $O_{R}=O_{S}\otimes I_{E}$ and more general than those of
other formalisms. This is why we can find the moving preferred basis in a
general case containing EID as particular case. Anyhow the analyticity
conditions must also be satisfied. In the case of EID we can substitute
$O_{R}$ by $O_{S}$ and $\rho_{R}(t)$ by $\rho_{S}(t).$}. In fact, in the next
subsection we will only consider the generic mean value $(\rho_{R}(t)|O_{R})$
for three paradigmatic model below. Model 1 with just one pole and the Khalfin
effect. Model 2 with two poles and Model 3 with $N$ poles.

\subsection{Model 1. One pole and the Khalfin term:}

It can be proved (cf. (\cite{JPA}) eq. (67)) that the evolution equation of
the mean value $(\rho(t)|O)$ is
\begin{align}
\langle O_{R}\rangle_{\rho(t)}  &  =(\rho(t)|O_{R})=(\rho_{R}(t)|O_{R}%
)\nonumber\\
&  =\int_{0}^{\infty}\rho^{\ast}(\omega)O(\omega)\,d\omega+\int_{0}^{\infty
}\int_{0}^{\infty}\rho^{\ast}(\omega,\omega^{\prime})O(\omega,\omega^{\prime
})\,e^{i\frac{\omega-\omega^{\prime}}{\hbar}t}\,d\omega d\omega^{\prime}
\label{C}%
\end{align}
i.e. this real mean value reads%

\begin{equation}
(\rho_{R}(t)|O_{R})=\int d\omega(\rho(0)|\Phi_{\omega})(\widetilde
{\Phi_{\omega}}|O_{R})+\int_{0}d\omega\int_{0}d\omega^{\prime}e^{\frac
{i}{\hbar}(\omega-\omega^{\prime})t}(\rho_{R}(0)|\Phi_{\omega\omega^{\prime}%
})(\widetilde{\Phi_{\omega\omega^{\prime}}}|O_{R})
\end{equation}

Where $O_{\omega}=(\widetilde{\Phi_{\omega}}|O_{R}),$ $O_{\omega\omega
^{\prime}}=(\widetilde{\Phi_{\omega\omega^{\prime}}}|O_{R}),$ $\rho_{\omega
}=(\rho_{R}(0)|\Phi_{\omega}),$ $\rho_{\omega\omega^{\prime}\text{ \ }}%
=(\rho_{R}(0)|\Phi_{\omega\omega^{\prime}}).$ These $\Phi$ vectors are defined
in eqs. (\ref{53'}) and (\ref{54}). Then, if we endow the functions with
analytical properties of subsection C and there is just one pole $z_{0}$ in
the lower halfplane, we can prove (\cite{JPA} eq. (70)) that%
\begin{align}
(\rho_{R}(t)|O_{R})  &  =\int d\omega(\rho_{R}(0)|\Phi_{\omega})(\widetilde
{\Phi_{\omega}}|O_{R})\nonumber\\
&  +e^{\frac{i}{\hbar}(z_{0}^{\ast}-z_{0})t}(\rho_{R}(0)|\Phi_{00}%
)(\widetilde{\Phi_{00}}|O_{R})\nonumber\\
&  +\int_{\Gamma}dz^{\prime}e^{\frac{i}{\hbar}(z_{0}^{\ast}-z^{\prime})t}%
(\rho_{R}(0)|\Phi_{0z^{\prime}})(\widetilde{\Phi_{0z^{\prime}}}|O_{R}%
)\nonumber\\
&  +\int_{\Gamma^{\ast}}dze^{\frac{i}{\hbar}(z-z_{0})t}(\rho_{R}(0)|\Phi
_{0z})(\widetilde{\Phi_{0z}}|O_{R})\nonumber\\
&  +\int_{\Gamma^{\ast}}dz\int_{\Gamma}dz^{\prime}e^{\frac{i}{\hbar
}(z-z^{\prime})t}(\rho_{R}(0)|\Phi_{zz^{\prime}})(\widetilde{\Phi_{zz^{\prime
}}}|O_{R}) \label{70}%
\end{align}
where $z_{0}$ $=\omega_{0}-\frac{i}{2}\gamma_{0},$ $\gamma_{0}>0$ and where
$|\Phi_{z}),(\widetilde{\Phi_{z}}|,|\Phi_{zz^{\prime}}),$ and $(\widetilde
{\Phi_{zz^{\prime}}}|$ are the analytical continuation in the lower half-plane
of (see (\cite{JPA} eq. (54))%
\begin{align}
|\Phi_{\omega})  &  =|\omega^{+}\rangle\langle\omega^{+}|\nonumber\\
\widetilde{(\Phi_{\omega}|}  &  =\widetilde{(\omega|},\text{ }|\Phi
_{\omega\omega^{\prime}})=|\omega^{+}\rangle\langle\omega^{+\prime}|,
\label{53'}%
\end{align}
and%
\begin{equation}
(\widetilde{\Phi_{\omega\omega^{\prime}}}|=\int d\varepsilon\lbrack
\langle\omega^{+}|\varepsilon\rangle\langle\varepsilon|\omega^{\prime+}%
\rangle-\delta(\omega-\varepsilon)\delta(\omega^{\prime}-\varepsilon
)](\widetilde{\varepsilon}|+\int d\varepsilon\int d\varepsilon^{\prime}%
\langle\omega^{+}|\varepsilon\rangle\langle\varepsilon^{\prime}|\omega
^{\prime+}\rangle\widetilde{(\varepsilon,\varepsilon^{\prime}}| \label{54}%
\end{equation}
and where $z_{0}$ is the simple pole of Figure 1 in the lower half-plane.
$|\Phi_{z}),(\widetilde{\Phi_{z}}|,|\Phi_{zz^{\prime}}),$ and $(\widetilde
{\Phi_{zz^{\prime}}}|$ can be defined as in the case of eq. (\ref{L.1}) and
(\ref{L.2}). The $|\Phi_{z}),(\widetilde{\Phi_{z}}|,|\Phi_{zz^{\prime}}),$ and
$(\widetilde{\Phi_{zz^{\prime}}}|$ can also be defined as a simple
generalization of the vectors $|f_{0}\rangle,$ $\langle\widetilde{f_{0}}|,$
$|f_{z}\rangle,$ and $\langle\widetilde{f_{z}}|$ (\cite{JPA}. eq. (42)). Then
the eqs. (\ref{53'}) and (\ref{54}) allow us to compute the limits
(\ref{INT-01}) and (\ref{INT-02'}) for any $\rho_{R}(0).$

Therefore we can conclude than the last four terms of equation (\ref{70})
\ vanish with characteristic times%
\begin{equation}
\frac{\hbar}{\gamma_{0}};\frac{2\hbar}{\gamma_{0}};\text{ }\frac{2\hbar
}{\gamma_{0}};\infty\label{TC}%
\end{equation}
respectively. Let us observe that

i-. The vanishing of the second, third, and fourth therms of eq. (\ref{70})
are \textit{exponential decaying}. This will also be the case in more
complicated models with many poles.

ii.- The $\infty$ means that the evolution of the last term of this equation
corresponds to a polynomial decaying in $t$, \ i. e. to the \textit{Khalfin
evolution}. This is a very weak effect detected in 2006 \cite{KhalfinEx}. If
there is a finite number of poles the Khalfin term corresponds to the integral
along the curve $\Gamma$ and contains the contribution of all the poles placed
bellow $\Gamma$\footnote{If the there is an infinite set of poles at $z_{i},$
with imaginary part $-\frac{1}{2}\gamma_{i}$ such that $lim_{i\rightarrow
\infty}\gamma_{i}=\infty$, then we can choose a curve $\Gamma_{j\text{ }}%
$below the poles a $\gamma_{1},\gamma_{2},...\gamma_{j}$. Then the integral
along the curve $\Gamma_{j}$ contains the effect of the poles $\gamma
_{j+1},\gamma_{j+2},...$ Thus we can choose the curve $\Gamma_{j\text{ }}$in
such a way hat the decaying times corresponding to these poles, $t_{j+n}%
=\hbar/\gamma_{j+n}$ would be so small that can be neglected.}. A closed
system model for Khalfin effect can be found in \cite{K1}, section 6, and an
EID-like model in \cite{K2}, section 5.

Now for times $t>$ $t_{D}=\frac{\hbar}{\gamma_{0}}$, eq. (\ref{70}) reads%
\begin{equation}
(\rho_{R}(t)|O_{R})=\int d\omega(\rho(0)|\Phi_{\omega})(\widetilde
{\Phi_{\omega}}|O)+\int_{\Gamma^{\ast}}dz\int_{\Gamma}dz^{\prime}e^{\frac
{i}{\hbar}(z-z^{\prime})t}(\rho(0)|\Phi_{zz^{\prime}})(\widetilde
{\Phi_{zz^{\prime}}}|O) \label{700}%
\end{equation}
since for $t>$ $t_{D}=\frac{\hbar}{\gamma_{0}}$ the poles term has
vanished\footnote{Since $t_{D}$ is just an order of magnitude we consider that
the three first imaginary parts of eqs. (\ref{TC}) and (\ref{TC'}) are
essentially equivalent.}.

Let us diagonalize $\rho_{R}(t)$ as \footnote{Here, for the sake of
simplicity, we will use sum instead of integral, as below in all cases of
diagonalization. Moreover, in many cases, the $O_{R}$ or the initial
conditions may just choose a discrete basis (see below).}%
\begin{equation}
\rho_{R}(t)=\sum_{i}\rho_{i}(t)|i(t)\rangle\langle i(t)| \label{asterisco}%
\end{equation}
where $\{|i(t)\rangle\}$ is the moving eigenbasis of $\rho_{R}(t)$.

Then let us define a state $(\rho_{P}(t)|,$ the \textit{preferred state}, such
that, \textit{for all times,} it would be%
\begin{equation}
(\rho_{P}(t)|O_{R})=\int d\omega(\rho(0)|\Phi_{\omega})(\widetilde
{\Phi_{\omega}}|O)+\int_{\Gamma^{\ast}}dz\int_{\Gamma}dz^{\prime}e^{\frac
{i}{\hbar}(z-z^{\prime})t}(\rho(0)|\Phi_{zz^{\prime}})(\widetilde
{\Phi_{zz^{\prime}}}|O) \label{asterisco'}%
\end{equation}
So $\rho_{P}(t)$ is a state that evolves in a model with no poles and with
only the Khalfin term . These evolutions exist and can be found using an
adequate interaction \footnote{All these formulas are confirmed by the
coincidence of results with other methods: e.g. those used to study a
$^{208}Pb(2d_{5/2})$ proton state in a Woods-Saxon potential (see \cite{JPA}
Figure 3).}. So we can plot $F(t)=(\rho_{R}(t)|O_{R})-(\rho_{P\ast}|O_{R})$
and $F_{Khalfin}(t)=(\rho_{P}(t)|O_{R})-(\rho_{P\ast}|O_{R})$ in figure 2.

\begin{figure}[t]
\centerline{\scalebox{0.7}{\includegraphics{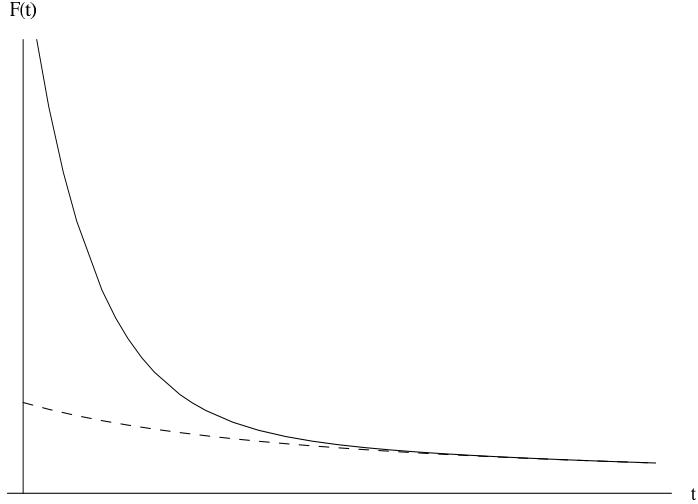}}} \vspace
*{0.cm}\caption{Evolution of $F(t)$ (solid line), $F_{Khalfin}(t)$ (dashed
line) and their coincidence limit at $t_{D}$.}%
\label{fig 2}%
\end{figure}

It is quite clear that for $t>t_{D}$ $\rho_{R}(t)\neq$ $\rho_{P}(t)$ while for
$t<t_{D}$ $\rho_{R}(t)=$ $\rho_{P}(t)$ and that for $t\rightarrow t_{D}$
$\rho_{R}(t)\rightarrow$ $\rho_{P}(t)$ and also all their derivatives.

The eigen states of the $\rho_{P}(t)$ are those that we will choose for the
moving decoherence basis. In fact, diagonalizing $\rho_{P}(t)$ we have
\begin{equation}
\rho_{P}(t)=\sum_{j}\rho_{j}(t)\widetilde{|j(t)\rangle}\widetilde{\langle
j(t)|} \label{asterisco'-b}%
\end{equation}
and when $t\rightarrow$ $t_{D}=\frac{\hbar}{\gamma_{0}}$ we have that
$\rho(t)\rightarrow\rho_{P}(t)$ so from eqs. (\ref{asterisco}) and
(\ref{asterisco'-b}) we see that the eigenbasis of $\rho(t)$ and $\rho_{P}(t)$
also converge
\begin{equation}
\left\{  |i(t)\rangle\right\}  \rightarrow\{\widetilde{|j(t)\rangle}\}
\label{asterisco'-c}%
\end{equation}
Namely the basis $\left\{  |i(t)\rangle\right\}  $ converge to $\{\widetilde
{|j(t)\rangle}\}$ and therefore $\rho_{R}(t)$ becomes diagonal in
$\{\widetilde{|j(t)\rangle}\}$. Thus $\{\widetilde{|j(t)\rangle}\}$ is our
definition for the \textit{moving preferred basis}. Since $\rho_{R}(t)$
becomes diagonal in the just defined preferred basis $\{\widetilde
{|j(t)\rangle}\}$ when $t\rightarrow t_{D}$ and $t_{D}=\frac{\hbar}{\gamma
_{0}}$ is the definition of decoherence time. In this model the relaxation
time $t_{R}$ is the corresponding to the Khalfin term, i.e. an extremely long
time so%
\begin{equation}
t_{D}\ll t_{R} \label{nuevo-0}%
\end{equation}

\subsection{Model 2: Two poles are considered and the Khalfin term is
neglected.}

The Khalfin term is so small (see \cite{KhalfinEx}) that can be neglected in
most of the experimental cases. So let us consider the case of two poles
$z_{0}$ and $z_{1}$ (and no relevant Khalfin term) where eq. (\ref{70}) reads:%

\[
(\rho_{R}(t)|O_{R})=\int d\omega(\rho_{R}(0)|\Phi_{\omega})(\widetilde
{\Phi_{\omega}}|O_{R})+e^{\frac{i}{\hbar}(z_{0}^{\ast}-z_{0})t}(\rho
_{R}(0)|\Phi_{00})(\widetilde{\Phi_{00}}|O_{R})+
\]%
\begin{equation}
e^{\frac{i}{\hbar}(z_{1}^{\ast}-z_{0})t}(\rho_{R}(0)|\Phi_{10})(\widetilde
{\Phi_{10}}|O_{R})+e^{\frac{i}{\hbar}(z_{0}^{\ast}-z_{1})t}(\rho_{R}%
(0)|\Phi_{01})(\widetilde{\Phi_{01}}|O_{R})+e^{\frac{i}{\hbar}(z_{1}^{\ast
}-z_{0})t}(\rho_{R}(0)|\Phi_{11})(\widetilde{\Phi_{11}}|O_{R}) \label{70'}%
\end{equation}
where $z_{0}$ $=\omega_{0}-\frac{i}{2}\gamma_{0},$ $\gamma_{0}>0$ , $z_{1}$
$=\omega_{1}-\frac{i}{2}\gamma_{1},$ $\gamma_{1}>0,$ and we will also consider
that $\gamma_{0}\ll\gamma_{1}$ (see \cite{Manoloetal} section 3, for details).
Then the four characteristic times (\ref{TC}) now read%
\begin{equation}
\frac{\hbar}{\gamma_{0}};\frac{\hbar}{\gamma_{1}+\gamma_{0}};\frac{\hbar
}{\gamma_{1}+\gamma_{0}}\approx\frac{\hbar}{\gamma_{1}} \label{TC'}%
\end{equation}
Now for times $t>$ $t_{D}=\frac{\hbar}{\gamma_{1}}$, eq. (\ref{700}) reads%
\[
(\rho_{R}(t)|O_{R})=\int d\omega(\rho_{R}(0)|\Phi_{\omega})(\widetilde
{\Phi_{\omega}}|O_{R})+e^{\frac{i}{\hbar}(z_{0}^{\ast}-z_{0})t}(\rho
_{R}(0)|\Phi_{00})(\widetilde{\Phi_{00}}|O_{R})
\]
and we can define a state $(\rho_{P}(t)|$ such that, for \textit{all times},
it would be%
\begin{equation}
(\rho_{P}(t)|O_{R})=\int d\omega(\rho_{R}(0)|\Phi_{\omega})(\widetilde
{\Phi_{\omega}}|O_{R})+e^{\frac{i}{\hbar}(z_{0}^{\ast}-z_{0})t}(\rho
_{R}(0)|\Phi_{00})(\widetilde{\Phi_{00}}|O_{R}) \label{asterisco''}%
\end{equation}
Repeating the reasoning of eqs. (\ref{700}) to (\ref{asterisco'-c}) we can see
that, diagonalizing this last equation, we obtain the moving preferred basis.
Then in this case we see that the \ relaxation is obtained by an exponential
dumping (not a Khalfin term) and
\begin{equation}
t_{R}=\frac{\hbar}{\gamma_{0}}\gg t_{D}=\frac{\hbar}{\gamma_{1}} \label{nuevo}%
\end{equation}
Again, in this case when $t\rightarrow$ $t_{D}=\frac{\hbar}{\gamma_{0}}$ we
have that $\rho_{R}(t)\rightarrow\rho_{P}(t)$ so once more we reach eq.
(\ref{asterisco'-c}). Namely $\rho(t)$ becomes diagonal in the moving
preferred basis in a time $t_{D}$.

\subsubsection{Remarks}

Before considering the many poles case let us make some general remarks.

i.- Let us observe that some $(\widetilde{\Phi_{\omega}}|O_{R}),$
$(\widetilde{\Phi_{0z^{\prime}}}|O_{R}),$ $(\widetilde{\Phi_{0z}}|O_{R})$ and
$(\widetilde{\Phi_{zz^{\prime}}}|O_{R})$ may be zero, depending in the
observable $O_{R},$ so, in the case of many poles, may be some poles can be
detected by $O_{R}$ and others may not be detected and disappear from the
formulae (see Appendix).

This also is the cases for the initial conditions: $(\rho_{R}(0)|\Phi_{\omega
}),$ $(\rho_{R}(0)|\Phi_{0z^{\prime}}),$ $(\rho_{R}(0)|\Phi_{0z}),$ and
$(\rho_{R}(0)|\Phi_{zz^{\prime}})$ may be zero. But also the $O_{R}$ or the
$\rho_{R}(0)$ may create some poles. So some poles may be eliminated or
created by the observables or the initial conditions while others may be
retained. But in general we will choose $O_{R}$ and $\rho_{R}(0)$ in such a
way that they would neither create or eliminate poles.

ii.- From what we have learned in both models (see eqs. (\ref{nuevo-0}) and
(\ref{nuevo})) we always have%
\begin{equation}
t_{D}<t_{R} \label{ECR-01}%
\end{equation}

\subsection{Model 3: The $N$ poles case.}

Let us now sketch the case of a system with $N$ poles located at $z_{i}%
=\omega_{i}^{\prime}-i\gamma_{i}.$ These poles are the ones that remain after
$O_{R}$ and $\rho_{R}(0)$ have eliminated (or created) some poles (see remark
i). In this case it is easy to see that eq. (\ref{70'}) (with no Khalfin term)
becomes:%
\begin{equation}
(\rho_{R}(t)|O_{R})=(\rho_{R\ast}|O_{R})+\sum_{i}a_{i}\exp\left(  -\frac
{i}{\hbar}\gamma_{i}t\right)  \label{Suma}%
\end{equation}
where $(\rho_{R\ast}|O_{R})$ is the final equilibrium value of $(\rho
_{R}(t)|O_{R}).$ In the most general case the $z_{i}$ will be placed either at
random or not. Anyhow in both cases they can be ordered as%
\[
\gamma_{0}\leq\gamma_{1}\leq\gamma_{2}\leq...
\]
So in the case of $3$ poles, we can plot $F(t)=(\rho_{R}(t)|O_{R}%
)-(\rho_{R\ast}|O_{R})=\sum_{i=0}^{2}a_{i}\exp\left(  -\frac{i}{\hbar}%
\gamma_{i}t\right)  $, $F_{\gamma_{0},\gamma_{1}}(t)=\sum_{i=0}^{1}a_{i}%
\exp\left(  -\frac{i}{\hbar}\gamma_{i}t\right)  $ and $F_{\gamma_{0}}%
(t)=a_{0}\exp\left(  -\frac{i}{\hbar}\gamma_{0}t\right)  $ in figure 3.

\begin{figure}[t]
\centerline{\scalebox{0.7}{\includegraphics{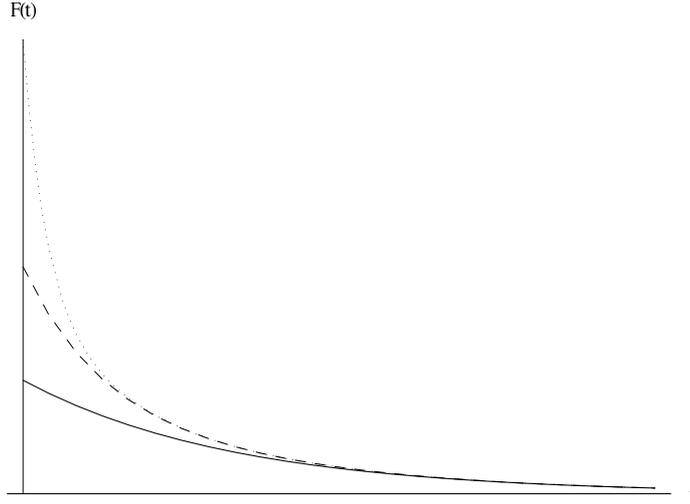}}} \vspace
*{0.cm}\caption{Evolution of $F(t)$ (solid line), $F_{\gamma_{0},\gamma_{1}%
}(t)$ (dashed line), $F_{\gamma_{0}}(t)$ (dot line)\ and their coincidence
limit at $t_{D}$. We can see the dominant components in different periods of
time.}%
\label{fig 3}%
\end{figure}

Then if $\gamma_{0}<\gamma_{1}$ it is quite clear that the relaxation time is
$t_{R}=\frac{\hbar}{\gamma_{0}}.$ So the relaxation time is defined with no ambiguity.

This is not the case for the decoherence time. Really each pole $z_{i}$
defines a decaying mode with characteristic time%
\[
t_{i}=\frac{\hbar}{\gamma_{i}}%
\]
These poles contain the essence of the decaying phenomenon and the definition
of the decoherence time depends on their distribution and other data like the
initial condition. Precisely

i.- For a completely random distribution clearly the best choice is%
\[
t_{D}=\frac{\hbar}{\gamma_{1}}%
\]
Then
\begin{equation}
(\rho_{P}(t)|O_{R})=(\rho_{R\ast}|O_{R})+a_{0}\exp\left(  -\frac{i}{\hbar
}\gamma_{0}t\right)  \label{def}%
\end{equation}
and the moving preferred basis is $\{\widetilde{|j(t)\rangle}\},$ i.e. the
basis that diagonalizes $\rho_{P}(t).$

ii.- But for other kinds of distributions, if the distribution of poles obey
certain law or have some patterns, we may chose something like%
\begin{equation}
t_{D}=\frac{\hbar}{f(\gamma_{0},\gamma_{1},\gamma_{2},...)} \label{34'}%
\end{equation}
as we will see in the example in the next section.

The choice of $f(\gamma_{0},\gamma_{1},\gamma_{2},...)$ is based on the
\textit{initial conditions} and usually also introduces the concept of
\textit{macroscopicity}\footnote{See eq. (\ref{CI-42}) and variable $L_{0}.$}.
Example of this choice are the Omn\`{e}s model and the toy models in the final
remake i of the next section.

Once the decoherence time $t_{D\text{ }}$is chosen the definition (\ref{def})
changes to%
\begin{equation}
(\rho_{P}(t)|O_{R})=(\rho_{R\ast}|O_{R})+\sum_{j=1}^{M}a_{j}\exp\left(
-\frac{i}{\hbar}\gamma_{j}t\right)  \label{EV}%
\end{equation}
where the sum contains all the terms such that
\[
\gamma_{i}\leq\frac{\hbar}{t_{D}}%
\]
Then the poles in evolution (\ref{EV}) are those that produces the slowest
decaying modes that we will call the \textit{p-relevant poles, }i. e. those
that have influence in the period $t_{\text{ }}>t_{D\text{ }}.$\textit{ }The
remaining poles such that
\[
\gamma_{i}>\frac{\hbar}{t_{D}}%
\]
that we will can the \textit{p-irrelevant poles}, and have no influence in the
period $t_{\text{ }}>t_{D\text{ }}$

Once the decoherence time is chosen the moving preferred basis is univocally
defined. It is $\{\widetilde{|j(t)\rangle}\}$, the basis that diagonalize
$\rho_{P}(t)$, a state that evolves only influenced by the p-relevant poles,
and such that $\rho_{P}(t)\rightarrow\rho_{R}(t)$ when $t\rightarrow t_{D}.$

$\{\widetilde{|j(t)\rangle}\}$ is our candidate for a general definition of
moving preferred basis.

\section{\label{Polos}The Omn\`{e}s or Lee-Friedrich model.}

\subsection{Omn\`{e}s formalism.}

Our more complete and simplest example of decoherence in open systems is the
Omn\`{e}s \textquotedblleft pendulum\textquotedblright\ (i. e. oscillator) in
a bath of oscillators, that we will compare with the poles theory in the
following subsections. In fact the Omn\`{e}s model could be considered a poles
model if we retain the poles and neglect the Khalfin term. Moreover in the
Omn\`{e}s philosophy the moving preferred basis must be related to some
\textquotedblleft collective variables\textquotedblright\ in such a way that
they would be experimentally accessible. In this case this variable is the
center of mass of the pendulum, i. e. the mean value of the position of a
coherent state. In \cite{Omnesazul} page 285 a one dimensional "pendulum" (the
system) in a bath of oscillators (the environment) is considered. The
Hamiltonian reads \footnote{This Hamiltonian is similar to the one of equation
(\ref{Hamil}) and equation (\ref{PRE}). In fact, in some stages of the
treatment Omn\`{e}s is forced go to the continuos spectrum. A complete
treatment of this continuous model can be found in \cite{PRE}.}%
\begin{equation}
H=\omega a^{\dagger}a+\sum_{k}\omega_{k}b_{k}^{\dagger}b_{k}+\sum_{k}%
(\lambda_{k}a^{\dagger}b_{k}+\lambda_{k}^{\ast}ab_{k}^{\dagger}) \label{1}%
\end{equation}
where $a^{\dagger}(a)$ is the creation (annihilation) operator for the system,
$b_{k}^{\dagger}(b_{k})$ are the creation (annihilation) operator for each
mode of the environment, $\omega$ and $\omega_{k}$ are the energies of the
system and each mode of the environment and $\lambda_{k}$ are the interaction coefficients.

Then let consider a state%
\[
|\psi(t)\rangle=a|\alpha_{1}(t)\rangle\prod_{k}|\beta_{k1}(t)\rangle
+b||\alpha_{2}(t)\rangle\prod_{k}|\beta_{k2}(t)\rangle
\]
where $|\alpha_{1}(0)\rangle,|\alpha_{2}(0)\rangle$ are \textit{coherent}
states for the "system" corresponding to the operator $a^{\dagger}$ and
$|\beta_{k1}(0)\rangle,$ $|\beta_{k2}(0)\rangle$ are a coherent state for the
environment corresponding to the operator $b_{k}^{\dagger}.$ Let the initial
condition be%
\begin{equation}
|\psi(0)\rangle=a|\alpha_{1}(0)\text{ \{}\beta_{k1}(0)=0\}\rangle+b|\alpha
_{2}(t),\{\beta_{k2}(0)=0\rangle\label{in}%
\end{equation}
Then
\begin{equation}
\rho_{R}(0)=Tr_{E}|\psi(0)\rangle\langle\psi(0)|\text{ and }\rho_{R}%
(t)=Tr_{E}|\psi(t)\rangle\langle\psi(t)| \label{tr}%
\end{equation}
Moreover it can be shown, under reasonable hypotheses and approximations (that
correspond to the elimination of the Khalfin terms, see below), that evolution
of the $|\alpha_{1}(t)\rangle,|\alpha_{2}(t)\rangle$ is given by
\begin{equation}
\alpha(t)=\alpha(0)\exp[-i(\omega+\delta\omega)t-\gamma t]+\text{small
fluctuations} \label{RT-01}%
\end{equation}
where $\delta\omega$ is a shift and $\gamma$\ a dumping coefficient that
produces that the system arrives at a state of equilibrium at $t_{R}%
=\hbar/\gamma$, the \textit{relaxation time} of the system, (the small
fluctuations are usually neglected)

In the next subsections using the concepts of the previous section we will
prove that the Omn\`{e}s model is a particular case of our general scheme. Let
us now consider the condition of \textit{experimentally accessibility. }In
fact, in the model under consideration, the initial states corresponds to the
linear combination of two coherent, macroscopically different states
$|\alpha_{1}(0)\rangle,|\alpha_{2}(0)\rangle$ that evolve to $|\alpha
_{1}(t)\rangle,|\alpha_{2}(t)\rangle$ .

Now the diagonal part of $\rho_{R}(t)$ reads
\[
\rho_{R}^{(D)}(t)=|a|^{2}|\alpha_{1}(t)\rangle\langle\alpha_{1}(t)|+|b|^{2}%
|\alpha_{2}(t)\rangle\langle\alpha_{2}(t)|
\]
and, it can easily be shown \cite{Omnesazul} that, with the choice of initial
conditions of eqs. (\ref{CI-17}) and (\ref{CI-18}), that the non diagonal part
of $\rho_{R}(t)$ is%

\[
\rho_{R}^{(ND)}(t)=(ab^{\ast}|\alpha_{1}(0)\rangle\langle\alpha_{2}%
(0)|+ba^{\ast}|\alpha_{2}(0)\rangle\langle\alpha_{1}(0)|)\exp\left[  -\frac
{1}{4}\frac{m\omega^{2}}{\hbar}(x_{1}(0)-x_{2}(0))^{2}(1-e^{-2\gamma
t})\right]
\]
Then if $t\ll t_{R}=\frac{\hbar}{\gamma}$ (that will be the case if
$L_{0}=|x_{2}(0)-x_{1}(0)|$ is very big) we have%
\begin{equation}
\rho_{R}^{(ND)}(t)\sim(ab^{\ast}|\alpha_{1}(0)\rangle\langle\alpha
_{2}(0)|+ba^{\ast}|\alpha_{2}(0)\rangle\langle\alpha_{1}(0)|)\exp\left[
-\frac{1}{4}\frac{m\omega^{2}}{\hbar}(x_{1}(0)-x_{2}(0))^{2}(1-1+2\frac
{t}{t_{R}}+...)\right]  \label{OM}%
\end{equation}
where $x_{1}(0),x_{2}(0)$ are the initial mean value of the position of the
two coherent states $|\alpha_{1}(t)\rangle,|\alpha_{2}(t)\rangle$. This
decaying structure is obviously produced by the combination of the initial
states and the particular evolution of the system according to the discussion
in the final part of the last section. Then, since $\rho_{R}^{(ND)}%
(t)\rightarrow0$ when $t\rightarrow\infty,$ $\rho_{R}(t)$ decoheres in the
decoherence basis \{$|\alpha_{1}(t)\rangle,|\alpha_{2}(t)\rangle\}$, which is
the moving preferred basis, and the decoherence time of the system is%

\begin{equation}
t_{D}(L_{0})\sim\lbrack m\omega^{2}(x_{1}(0)-x_{2}(0))^{2}]^{-1}t_{R}
\label{131'}%
\end{equation}
where $L_{0}=$%
$\vert$%
$x_{1}(0)-x_{2}(0)|.$

In the next subsection we will see that we are dealing with a many poles model
where the effect of decoherence is produced by these poles and the particular
coherent states initial conditions, which produce a "new collective pole mode"
with $\gamma^{\prime}=$ $-\frac{1}{2}\frac{m\omega^{2}}{\hbar}(x_{1}%
(0)-x_{2}(0))^{2}$

In the case of the "pendulum" the moving preferred basis \{$|\alpha
_{1}(t)\rangle,|\alpha_{2}(t)\rangle\}$ is clear experimentally accessible
since, in principle, the mean value of the position $x_{1}(t),x_{2}(t),$ of
the two coherent states $|\alpha_{1}(t)\rangle,|\alpha_{2}(t)\rangle$ can be
measured and the $x_{1}(0)$ and $x_{2}(0)$ turn out to be two "collective
variables" (since they are mean values). In fact, in this formalism, the main
characteristic of the moving preferred basis is to be related to the
"collective variables". Moreover the decoherence time $t_{D}$ \ depends on the
initial distance $L_{0}=|x_{1}(0)-x_{2}(0)|$ so we can have different
decoherence times depending on the initial conditions.

Let us now consider that
\[
\langle\alpha_{1}(t)|\alpha_{2}(t)\rangle=\exp\left[  -\frac{|\alpha
_{1}-\alpha_{2}|^{2}}{2}+i\frac{\Phi}{2}\right]  ,\text{ \ \ \ }%
\Phi=\operatorname{Im}(\alpha_{1}\alpha_{2}^{\ast}-\alpha_{1}^{\ast}\alpha
_{2})
\]
where
\[
|\alpha_{1}-\alpha_{2}|=(2m\hbar\omega)^{-\frac{1}{2}}[m^{2}\omega^{2}%
(x_{1}(t)-x_{2}(t))^{2}+(p_{1}(t)-p_{2}(t))^{2}]^{\frac{1}{2}}%
\]
So:

i.- $\langle\alpha_{1}(t)|\alpha_{1}(t)\rangle=1$ even if in general
$\langle\alpha_{1}(t)|\alpha_{2}(t)\rangle\neq0$.

ii.- When $(x_{1}(0)-x_{2}(0))^{2}\rightarrow\infty$ or $(p_{1}(t)-p_{2}%
(t))^{2}\rightarrow\infty$ we have $\langle\alpha_{1}(t)|\alpha_{2}%
(t)\rangle\rightarrow0.$

Thus when the distance between the two centers of the coherent states is very
big we have a small $t_{D}$ and the basis \{$|\alpha_{1}(t)\rangle,|\alpha
_{2}(t)\rangle\}$would be almost orthonormal. These are the main
characteristics of the experimental accessible decoherence basis of Omn\`{e}s.

But it is important to insist that, generally, $\left\{  \left\vert \alpha
_{i}(t)\right\rangle \right\}  $ is only a \textit{non-orthonormal moving
preferred basis}, that we can approximately suppose orthonormal only in the
macroscopic case, that is to say, when $x_{1}(0)$ and $x_{2}(0)$ are far apart.

In conclusion, in this macroscopic case $\left\{  \left\vert \alpha
_{i}(t)\right\rangle \right\}  $ becomes a orthonormal \textit{ moving
preferred basis }where $\rho_{R}(t)$ becomes diagonal in a very small time.
This will be the case of the decoherence basis in \cite{OmnesRojo}, chapter
17, and in a many examples that we can find in the bibliography
(\cite{Paz-Zurek}, \cite{Paz-Habib-Zurek}, \cite{Ex}). Without this
macroscopic property it is difficult to find any trace of a Boolean logic in
the moving decoherence basis context of the general case or in this section.
In fact, Omn\`{e}s obtains the Boolean logic in a complete different way (see
chapter 6 of \cite{Omnesazul}).

Anyhow in this particular model the moving preferred basis has a perfect
example for the macroscopic case\footnote{This is the basis that it is
equivalent to the one introduced in section II F, as we will see.}. Let us now
present the relation of this formalism with the poles theory.

\subsection{\label{PolosH}Poles of the Lee-Friedrich model. The relaxation
time.}

Particular important models can be studied, like the one in \cite{PRE}, with Hamiltonian%

\begin{equation}
H=\omega_{0}a^{\dagger}a+\int\omega_{\mathbf{k}}b_{\mathbf{k}}^{\dagger
}b_{\mathbf{k}}d\mathbf{k}+\int\lambda_{\mathit{k}}(a^{\dagger}b_{\mathbf{k}%
}+ab_{\mathbf{k}}^{\dagger})d\mathbf{k} \label{PRE}%
\end{equation}

i.e. a continuous version of (\ref{1}). In this continuous version we are
forced to endow the scalar $(\rho_{R}(t)|O_{R})$ with the some analyticity
conditions. Precisely function $\lambda_{\mathit{k}}$ (where $k=\omega
_{k}=|\mathbf{k|)}$ is chosen in such a way that%
\begin{equation}
\eta_{\pm}(\omega_{k})=\omega_{k}-\omega_{0}-\int\frac{d\mathbf{k}%
\lambda_{\mathit{k}}^{2}}{\omega_{k}-\omega_{k^{\prime}}\pm i0}
\label{cuac-00}%
\end{equation}
which does not vanish for $k\in\mathbb{R}^{+},$ and its analytic extension
$\eta_{+}(z)$ to the lower half plane only has a simple pole at $z_{0}$. This
fact will have influence on the poles of $(\rho_{R}(t)|O_{R})$ as in section
II and we know that the study of $(\rho_{R}(t)|O_{R})$ is the essential way to
understand the whole problem (see section I A).

The Hamiltonian (\ref{PRE}) is sometimes called the Lee-Friedrich Hamiltonian
and it is characterized by the fact that it contains different \textit{number
of modes sector} (number of particle sectors in QFT). In fact, $a^{\dagger}$
and $b_{\mathbf{k}}^{\dagger}$ are creation operators that allow to define
these numbers of mode sectors. e. g. the one mode sector will contain states
like $a^{\dagger}|0\rangle$ and $b_{\mathbf{k}}^{\dagger}|0\rangle$ (where
$a|0\rangle=$ $b_{\mathbf{k}}|0\rangle=0).$ Then the action of $\exp\left(
-\frac{i}{\hbar}Ht\right)  $ (or simple the one of $H)$ will conserve the
number of modes of this sector in just one mode, since in \ (\ref{PRE}) all
the destruction operators are preceded by a creation operator. This also is
the case for the $n-$mode sector. The Hamiltonian of the one mode sector, is
just the one of the so called Friedrich model i. e.
\begin{equation}
H_{F}=\omega_{0}\left\vert 1\right\rangle \left\langle 1\right\vert
+\int\omega_{k}\left\vert \omega\right\rangle \left\langle \omega\right\vert
d\omega+\int\left(  \lambda(\omega)\left\vert \omega\right\rangle \left\langle
1\right\vert +\lambda^{\ast}(\omega)\left\vert 1\right\rangle \left\langle
\omega\right\vert \right)  d\omega\label{cuac-04}%
\end{equation}
(expressed just in variable $\omega,$the one that it is analytically
continued$).$ As a consequence of the analyticity condition above this simple
Friedrich model just shows one resonance. In fact, th is resonance is produced
in $z_{0}$. Let $H_{F}$ be the Hamiltonian of the complex extended Friedrich
model, then\footnote{Only symbolically, since the poles belong to the scalar
$(\rho(t)|O),$ as in section II.}:%
\begin{equation}
H_{F}|z_{0}\rangle=z_{0}|z_{0}\rangle,\text{ \ \ \ }H_{F}|z\rangle
=z|z\rangle\label{cuac-05}%
\end{equation}
where $z_{0}=\omega_{0}+\delta\omega_{o}-i\gamma_{0}=\omega_{0}^{\prime
}-i\gamma_{0}$ is the only pole and $z\in\Gamma$.

The Lee-Friedrich model, describing the interaction between a quantum
oscillator and a scalar field, is extensively analyzed in the literature.
Generally, this model is studied by analyzing first the one excited mode
sector, i.e. the Friedrich model. Then, if we compute the pole, of this last
model, up to the second order in $\lambda_{k}$ we obtain that
\begin{equation}
z_{0}=\omega_{0}+\int\frac{d\mathbf{k}^{\prime}\lambda_{\mathit{k}^{\prime}%
}^{2}}{\omega_{0}-\omega_{k}+i0} \label{cuac-01}%
\end{equation}

So the pole (that will corresponds to the pole closest to the real axis in the
Lee-Friedrich model) can be calculated (see \cite{LCIB} eq. (84)). These
results coincide (mutatis mutandis) with the one of Omn\`{e}s book
\cite{Omnesazul}\ page 288, for the pole corresponding the relaxation time. In
fact:
\begin{equation}
\frac{1}{\omega_{0}-\omega^{\prime}+i0}=P\left(  \frac{1}{\omega_{0}%
-\omega^{\prime}}\right)  -i\pi\delta(\omega_{0}-\omega^{\prime})
\label{cuac-02}%
\end{equation}
where $P$ symbolizes the \textquotedblleft principal part\textquotedblright,
so%
\begin{equation}
z_{0}=\omega_{0}+P\int\frac{d\mathbf{k}^{\prime}\lambda_{\mathbf{k}^{\prime}%
}^{2}}{\omega_{0}-\omega_{k}}-i\pi\int d\mathbf{k}^{\prime}\lambda
_{\mathbf{k}^{\prime}}^{2}\delta(\omega_{0}-\omega_{k}) \label{cuac-03}%
\end{equation}
Then if $d\mathbf{k}=n(\omega)d\omega$ we have%
\begin{equation}
\delta\omega_{0}=P\int\frac{n(\omega^{\prime})d\omega^{\prime}\lambda
_{\omega^{\prime}}^{2}}{\omega_{0}-\omega^{\prime}},\text{ \ \ \ \ }\gamma
_{0}=\pi\int n(\omega^{\prime})d\omega^{\prime}\lambda_{\omega^{\prime}}%
^{2}\delta(\omega_{0}-\omega^{\prime}) \label{Omnes}%
\end{equation}
namely the results of \cite{Omnesazul}\ page 288, and the one contained in eq.
(\ref{RT-01}).:%
\begin{equation}
z_{0}=\left(  \omega_{0}+\delta\omega_{0}\right)  -i\gamma_{0}=\omega
_{0}^{\prime}-i\gamma_{0} \label{cuac}%
\end{equation}
So the Omn\`{e}s result for the decoherence time \textit{coincides}, as we
have already said, with the one obtained by the pole theory, so
\[
t_{R}=\frac{1}{\gamma_{0}}%
\]
in both frameworks.

\subsection{\label{PolosH copy(2)}Poles of the Lee-Friedrich model.}

Let us now consider the Lee-Friedrich Hamiltonian (\ref{PRE}) for the many
modes sector, e. g., as an example, for the three mode sector. Then we have
that\footnote{Only symbolically as we have already explained in a previous
footnote.}:%
\begin{equation}
H|z_{1},z_{2},z_{3}\rangle=(z_{1}+z_{2}+z_{3})|z_{1},z_{2},z_{3}%
\rangle\label{cuac-06}%
\end{equation}%
\begin{equation}
H|z_{1},z_{2},z_{0}\rangle=(z_{1}+z_{2}+z_{0})|z_{1},z_{2},z_{0}%
\rangle\label{cuac-07}%
\end{equation}%
\begin{equation}
H|z_{1},z_{0},z_{3}\rangle=(z_{1}+z_{0}+z_{3})|z_{1},z_{0},z_{3}%
\rangle\label{cuac-08}%
\end{equation}%
\begin{equation}
H|z_{0},z_{2},z_{3}\rangle=(z_{0}+z_{2}+z_{3})|z_{0},z_{2},z_{3}%
\rangle\label{cuac-09}%
\end{equation}%
\begin{equation}
H|z_{1},z_{0},z_{0}\rangle=(z_{1}+2z_{0})|z_{1},z_{0},z_{0}\rangle
\label{cuac-10}%
\end{equation}%
\begin{equation}
H|z_{0},z_{2},z_{0}\rangle=\left(  z_{2}+2z_{0}\right)  |z_{0},z_{2}%
,z_{0}\rangle\label{cuac-11}%
\end{equation}%
\begin{equation}
H|z_{0},z_{0},z_{3}\rangle=(z_{3}+2z_{0})|z_{0},z_{0},z_{3}\rangle
\label{cuac-12}%
\end{equation}%
\begin{equation}
H|z_{0},z_{0},z_{0}\rangle=3z_{0}|z_{0},z_{0},z_{0}\rangle\label{cuac-13}%
\end{equation}
where $z_{1},z_{2},z_{3}\epsilon\Gamma.$ So in the real complex plane the
spectrum of $H$ is

1.- From the eigenvalue $(z_{1}+z_{2}+z_{3})$ three points of the curve
$\Gamma$

2.- From the eigenvalue $(z_{1}+z_{2}+z_{0}),$ $(z_{1}+z_{0}+z_{3}),$
$(z_{0}+z_{2}+z_{3})$, a pole at $z_{0\text{ }}$and two points of the curve
$\Gamma$

3.- From the eigenvalue $(z_{1}+2z_{0}),$ $(z_{2}-2z_{0}),$ $(z_{3}+2z_{0})$ a
pole at $2z_{0},$ and one point of the curve $\Gamma$

4.- From the eigenvalue a pole at $3z_{0}$

See figure 4:

\begin{figure}[t]
\centerline{\scalebox{0.7}{\includegraphics{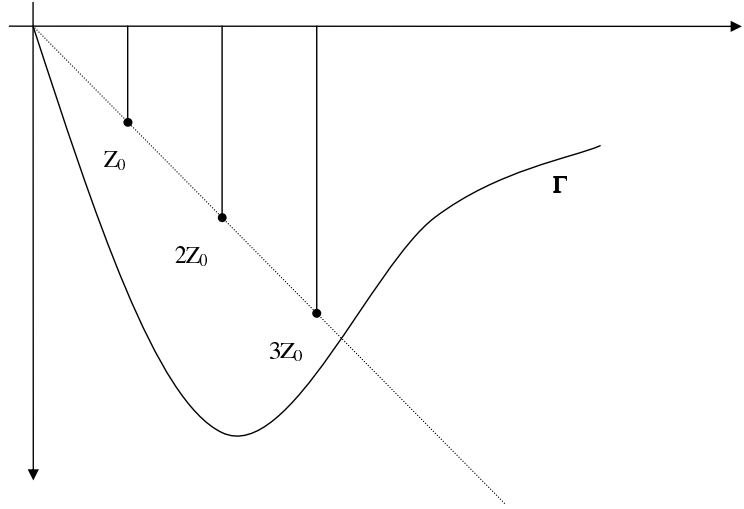}}} \vspace
*{0.cm}\caption{Complex contour on the lower complex energy plane for the
three modes model. The energy poles $z_{0},$ $2z_{0}$ $3z_{0}$ are assumed to
be simple.}%
\label{fig 4}%
\end{figure}

Of course in the general case $3\rightarrow n$ and as a consequence the
spectrum is $nz_{0}+$ the curves $\Gamma$, $n=0,1,2$. $3$,... in fact%
\begin{equation}
z_{n}=nz_{0}+\sum_{j}z_{j},\text{ }z_{j}\epsilon\Gamma\label{cuac-14}%
\end{equation}
Then if we neglect the Khalfin term, since it corresponds to extremely long
times, the $\Gamma$ disappears and we simply have%
\begin{equation}
z_{n}=nz_{0}=n(\omega_{0}^{\prime}-i\gamma_{0}) \label{cuac-15}%
\end{equation}
Then under this approximation the system has an effective (non Hermitian)
Hamiltonian%
\[
H_{eff}=\hbar z_{0}a_{0}^{\dagger}a_{0}=\hbar Nz_{0}%
\]
where $a_{0}^{\dagger},a_{0}$ are the creation and annihilation operators for
the mode corresponding to the pole $z_{0}$ and $N$ is the corresponding number
of poles operator. Now the Hamiltonian of the harmonic oscillator is%
\begin{equation}
H_{o}=\left(  N+\frac{1}{2}\right)  \hbar\omega\label{a}%
\end{equation}
Thus we see that in the no Khalfin terms approximation, and taking $\omega
_{0}^{\prime}=\omega$ (or the last equation) and if $n$ is very
large\footnote{Or, in the general case, since $\frac{\hbar\omega}{2}$ only
affects the real part of the pole and not the imaginary one that produces the
time scales.}
\begin{equation}
H_{eff}=H_{o}-i\frac{\gamma_{0}}{\hbar\omega}H_{o} \label{b}%
\end{equation}
\ So, in this approximation, the effective Lee-Friedrich Hamiltonian $H_{eff}$
simply is a (non Hermitian) version of $H_{o}$ with a dumping term
$\frac{\gamma_{0}}{\hbar\omega}H_{o}.$ Moreover the basis of $H_{eff}$ and
$H_{o}$ are the same one , i. e.$\{|n\rangle\}.$

\subsection{\label{PolosI}Poles dependency on the initial conditions}

\subsubsection{The amplitude of probability}

The probability amplitude that a pure state $\left\vert \varphi\right\rangle $
would be in the pure state $\left\vert \psi\right\rangle $ at time $t$ is:%
\begin{equation}
A(t)=\left\langle \psi|\varphi(t)\right\rangle \label{CI-01}%
\end{equation}
The most general linear superposition of the eigenvectors of $H_{eff}$, in
basis $\{|n\rangle\}$ is:%
\begin{equation}
\left\vert \psi\right\rangle =\sum_{n=0}^{N}a_{n}\left\vert n\right\rangle
\label{CI-02}%
\end{equation}
and the time evolution for $\left\vert \varphi\right\rangle $ must be:%
\begin{equation}
\left\vert \varphi(t)\right\rangle =\sum_{n=0}^{N}b_{n}\left\vert
n(t)\right\rangle \label{CI-03}%
\end{equation}

Then%

\begin{equation}
A(t)=\sum_{n,n^{\prime}=0}^{N}b_{n}a_{n}^{\ast}\left\langle n|n^{\prime
}(t)\right\rangle =\sum_{n,n^{\prime}=0}^{N}b_{n}a_{n^{\prime}}^{\ast
}A_{nn^{\prime}} \label{CI-04}%
\end{equation}
We can compute $A_{nn^{\prime}}=\left\langle n|n^{\prime}(t)\right\rangle
=\left\langle n|\exp\left(  -\frac{i}{\hbar}Ht\right)  |n^{\prime
}\right\rangle =\left\langle n|e^{-i\frac{z_{n}}{\hbar}t}|n^{\prime
}\right\rangle =e^{-i\frac{z_{n}}{\hbar}t}\delta_{nn^{\prime}}$, then%

\begin{equation}
A(t)=\sum_{n=0}^{N}b_{n}a_{n}^{\ast}e^{-i\frac{z_{n}}{\hbar}t} \label{CI-05}%
\end{equation}
where from eqs. (\ref{RT-01}) and (\ref{Omnes}), or eq. 4.47 of
\cite{Manoloetal} we have:
\begin{equation}
z_{n}=\omega_{n}^{\prime}+i\gamma_{n} \label{CI-06}%
\end{equation}
Then if we neglect the Khalfin term the "energy" levels are multiples of the
fundamental "energy" i. e.%
\begin{equation}
z_{n}=nz_{0} \label{CI-07}%
\end{equation}
where $z_{0}=\omega_{0}^{\prime}-i\gamma_{0}$ and the coefficients $a_{n}$ and
$b_{n}$\ depend in the initial conditions (according to eq. 4.26 of
\cite{Manoloetal}).

With the expression (\ref{cuac-15}) eq. (\ref{CI-05}) becomes%

\begin{equation}
A(t)=\sum_{n=0}^{N}b_{n}a_{n}^{\ast}e^{-i\frac{nz_{0}}{\hbar}t}=\sum_{n=0}%
^{N}b_{n}a_{n}^{\ast}\left(  e^{-i\frac{z_{0}}{\hbar}t}\right)  ^{n}
\label{CI-09}%
\end{equation}
The same recipe could be used in the fundamental scalar $(\rho_{R}(t)|O_{R})$
instead of $\langle\psi|\varphi(t)\rangle$ with similar results but with more
difficult calculations.

\subsubsection{Initial conditions and evolution}

As initial conditions, $|\alpha_{1}(0)\rangle,$ $|\alpha_{2}(0)\rangle,$ it is
possible to choose any linear combination of the elements $\left\{  \left\vert
n\right\rangle \right\}  $\ with $n=0,1...,\infty$. . So we can choose the
coherent states
\begin{equation}
\left\vert \lambda\right\rangle =e^{-\frac{\left\vert \lambda\right\vert ^{2}%
}{2}}\sum_{n=0}^{\infty}\frac{\lambda^{n}}{\sqrt{n!}}\left\vert n\right\rangle
\label{CI-09b}%
\end{equation}
But we can also choose as the boundary condition an approximated version where
the number modes is $N$ and we take $n=0,1...,N.$ namely an approximated
quasi-coherent states or quasi-Gaussian (that becomes a coherent state when
$N\rightarrow\infty$ as we will consider below$.$ Thus
\begin{equation}
\left\vert \lambda\right\rangle =\left(  \sum_{k=0}^{N}\frac{\left\vert
\lambda\right\vert ^{2k}}{k!}\right)  ^{-\frac{1}{2}}\sum_{n=0}^{N}%
\frac{\lambda^{n}}{\sqrt{n!}}\left\vert n\right\rangle \label{CI-09c}%
\end{equation}
Then let us choose the initial conditions as the sum of two quasi-Gaussian
functions, namely:%
\begin{equation}
\left\vert \Phi(0)\right\rangle =a\left\vert \alpha_{1}(0)\right\rangle
+b\left\vert \alpha_{2}(0)\right\rangle \label{CI-10}%
\end{equation}
where $\left\vert \alpha_{1}(0)\right\rangle $ and $\left\vert \alpha
_{2}(0)\right\rangle $ are quasi-coherent states, precisely%
\begin{equation}
\left\vert \alpha_{1}(0)\right\rangle =\left(  \sum_{k=0}^{N}\frac{\left\vert
\alpha_{1}(0)\right\vert ^{2k}}{k!}\right)  ^{-\frac{1}{2}}\sum_{n=0}^{N}%
\frac{\left(  \alpha_{1}(0)\right)  ^{n}}{\sqrt{n!}}\left\vert n\right\rangle
\label{CI-11}%
\end{equation}
and%
\begin{equation}
\left\vert \alpha_{2}(0)\right\rangle =\left(  \sum_{k=0}^{N}\frac{\left\vert
\alpha_{2}(0)\right\vert ^{2k}}{k!}\right)  ^{-\frac{1}{2}}\sum_{n=0}^{N}%
\frac{\left(  \alpha_{2}(0)\right)  ^{n}}{\sqrt{n!}}\left\vert n\right\rangle
\label{CI-12}%
\end{equation}
Thus the initial state is:%
\begin{align}
\rho_{0}  &  =\left\vert \Phi(0)\right\rangle \left\langle \Phi(0)\right\vert
=\left\vert a\right\vert ^{2}\left\vert \alpha_{1}(0)\right\rangle
\left\langle \alpha_{1}(0)\right\vert +ab^{\ast}\left\vert \alpha
_{1}(0)\right\rangle \left\langle \alpha_{2}(0)\right\vert \nonumber\\
&  +a^{\ast}b\left\vert \alpha_{2}(0)\right\rangle \left\langle \alpha
_{1}(0)\right\vert +\left\vert b\right\vert ^{2}\left\vert \alpha
_{2}(0)\right\rangle \left\langle \alpha_{2}(0)\right\vert \label{CI-13}%
\end{align}
Therefore the time evolved state is%
\begin{equation}
\rho(t)=\left\vert \Phi(t)\right\rangle \left\langle \Phi(t)\right\vert
=\rho_{D}(t)+\rho_{ND}(t) \label{CI-14}%
\end{equation}
where $\rho_{D}(t)$\ is the diagonal part (in the basis $\left\{  \left\vert
\alpha_{1}(0)\right\rangle ,\left\vert \alpha_{2}(0)\right\rangle \right\}  $)
of $\rho(t)$
\begin{equation}
\rho^{(D)}(t)=\left\vert a\right\vert ^{2}\left\vert \alpha_{1}%
(t)\right\rangle \left\langle \alpha_{1}(t)\right\vert +\left\vert
b\right\vert ^{2}\left\vert \alpha_{2}(t)\right\rangle \left\langle \alpha
_{2}(t)\right\vert \label{CI-15}%
\end{equation}
and $\rho^{(ND)}$\ is the non-diagonal part of $\rho(t)$%
\begin{equation}
\rho^{(ND)}(t)=ab^{\ast}\left\vert \alpha_{1}(t)\right\rangle \left\langle
\alpha_{2}(t)\right\vert +a^{\ast}b\left\vert \alpha_{2}(t)\right\rangle
\left\langle \alpha_{1}(t)\right\vert \label{CI-16}%
\end{equation}
We choose the two quasi-Gaussian (\ref{CI-11}) and (\ref{CI-12}) with center
at $p_{1,2}(0)=0$, (see \cite{Omnesazul} eq. (7.15) page 284) and%
\begin{equation}
\alpha_{1}(0)=\frac{m\omega}{\sqrt{2m\hbar^{2}\omega}}x_{1}(0) \label{CI-17}%
\end{equation}%
\begin{equation}
\alpha_{2}(0)=\frac{m\omega}{\sqrt{2m\hbar^{2}\omega}}x_{2}(0) \label{CI-18}%
\end{equation}
So $\alpha_{1}(0)$ and $\alpha_{2}(0)$ are real numbers.

Without loss of generality (since with a change of coordinates we can shift
$x_{1}(0)$ and $x_{2}(0))$ we can consider that the $\alpha_{1}(0)$ and
$\alpha_{2}(0)$ are both positive. For this reason we will interchange
$\alpha_{i}(0)$ and $\left\vert \alpha_{i}(0)\right\vert $ below.

\subsubsection{Components of the non-diagonal part of the state and the
macroscopic case}

Let us not consider $\rho^{(ND)}(t)$ in the basis of the initial condition
$\left\{  \left\vert \alpha_{1}(0)\right\rangle ,\left\vert \alpha
_{2}(0)\right\rangle \right\}  .$Then we have%
\begin{align}
\rho^{(ND)}(t)  &  =\rho_{11}^{(ND)}(t)\left\vert \alpha_{1}(0)\right\rangle
\left\langle \alpha_{1}(0)\right\vert +\rho_{12}^{(ND)}(t)\left\vert
\alpha_{1}(0)\right\rangle \left\langle \alpha_{2}(0)\right\vert \nonumber\\
&  +\rho_{21}^{(ND)}(t)\left\vert \alpha_{2}(0)\right\rangle \left\langle
\alpha_{1}(0)\right\vert +\rho_{22}^{(ND)}(t)\left\vert \alpha_{2}%
(0)\right\rangle \left\langle \alpha_{2}(0)\right\vert \label{CI-20}%
\end{align}

We will prove that for macroscopic initial conditions, i.e. when the peaks of
the two Gaussians are far from each other, the states $\left\{  \left\vert
\alpha_{1}(0)\right\rangle ,\left\vert \alpha_{2}(0)\right\rangle \right\}
$\ are quasi-orthogonal basis, i.e.%
\begin{equation}
\left\langle \alpha_{1}(0)|\alpha_{2}(0)\right\rangle \cong\left\langle
\alpha_{2}(0)|\alpha_{1}(0)\right\rangle \cong0 \label{CI-21}%
\end{equation}
and indeed this is the macroscopicity condition. In fat%
\begin{equation}
\left\langle \alpha_{1}(0)|\alpha_{2}(0)\right\rangle =\left(  \sum_{k=0}%
^{N}\frac{\left\vert \alpha_{1}(0)\right\vert ^{2k}}{k!}\right)  ^{-\frac
{1}{2}}\left(  \sum_{k=0}^{N}\frac{\left\vert \alpha_{2}(0)\right\vert ^{2k}%
}{k!}\right)  ^{-\frac{1}{2}}\sum_{n=0}^{N}\frac{\left(  \alpha_{1}%
(0)\alpha_{2}(0)\right)  ^{n}}{n!} \label{CI-21b}%
\end{equation}
So using the Cauchy product and the binomial theorem we have%
\begin{equation}
\left\langle \alpha_{1}(0)|\alpha_{2}(0)\right\rangle =\left(  \sum_{k=0}%
^{N}\frac{\left(  \left\vert \alpha_{1}(0)\right\vert ^{2}+\left\vert
\alpha_{2}(0)\right\vert ^{2}\right)  ^{k}}{k!}\right)  ^{-\frac{1}{2}}%
\sum_{n=0}^{N}\frac{\left(  \alpha_{1}(0)\alpha_{2}(0)\right)  ^{n}}{n!}
\label{CI-21c}%
\end{equation}
and again using the Cauchy product and the binomial theorem we have%
\begin{equation}
\left\langle \alpha_{1}(0)|\alpha_{2}(0)\right\rangle =\sum_{n=0}^{N}\frac
{1}{n!}\left(  -\frac{1}{2}\left(  \left\vert \alpha_{1}(0)\right\vert
^{2}+\left\vert \alpha_{2}(0)\right\vert ^{2}-2\alpha_{1}(0)\alpha
_{2}(0)\right)  \right)  ^{n} \label{CI-21d}%
\end{equation}
then%
\begin{equation}
\left\langle \alpha_{1}(0)|\alpha_{2}(0)\right\rangle =\sum_{n=0}^{N}\frac
{1}{n!}\left(  -\frac{\left(  \alpha_{1}(0)-\alpha_{2}(0)\right)  ^{2}}%
{2}\right)  ^{n} \label{CI-21e}%
\end{equation}
so for
$\vert$%
$\alpha_{1}(0)-\alpha_{2}(0)|\rightarrow\infty$ we have orthogonality as we
have promised to demonstrate.

Now we can consider the limit $N\rightarrow\infty$ . Thus the last scalar
product is equal to the truncated Taylor series of exponential function. Then
we may introduce $R_{N+1},$ the difference with the complete Taylor series,
and we obtain%
\begin{equation}
\left\langle \alpha_{1}(0)|\alpha_{2}(0)\right\rangle =e^{-\frac{\left(
\alpha_{1}(0)-\alpha_{2}(0)\right)  ^{2}}{2}}-R_{N+1} \label{CI-21f}%
\end{equation}
where $R_{N+1}$\ is a correction of order $N+1$
\begin{equation}
R_{N+1}=\frac{e^{\xi}}{\left(  N+1\right)  !}\left(  -\frac{\left(  \alpha
_{1}(0)-\alpha_{2}(0)\right)  ^{2}}{2}\right)  ^{N+1} \label{CI-21g}%
\end{equation}
with $\xi\in\left[  -\frac{\left(  \alpha_{1}(0)-\alpha_{2}(0)\right)  ^{2}%
}{2},0\right]  $, then we have%
\begin{equation}
R_{N+1}\leq\frac{1}{\left(  N+1\right)  !}\left(  -\frac{\left(  \alpha
_{1}(0)-\alpha_{2}(0)\right)  ^{2}}{2}\right)  ^{N+1} \label{CI-21h}%
\end{equation}
Thus we have two the orthogonality conditions:

\begin{enumerate}
\item To eliminate the first term of (\ref{CI-21f})%
\begin{equation}
e^{-\frac{\left(  \alpha_{1}(0)-\alpha_{2}(0)\right)  ^{2}}{2}}\ll1
\label{CI-21i}%
\end{equation}
i.e.
\begin{equation}
\left\vert \alpha_{1}(0)-\alpha_{2}(0)\right\vert \gg1 \label{CI-21j}%
\end{equation}

\item To eliminate the second term of (\ref{CI-21f}), $\left\vert
R_{N+1}\right\vert \ll1$, then%
\begin{equation}
\left\vert \frac{1}{\left(  N+1\right)  !}\left(  -\frac{\left(  \alpha
_{1}(0)-\alpha_{2}(0)\right)  ^{2}}{2}\right)  ^{N+1}\right\vert \ll1
\label{CI-21k}%
\end{equation}
i.e.%
\begin{equation}
\left\vert \alpha_{1}(0)-\alpha_{2}(0)\right\vert \ll\left[  2\left(
N+1\right)  !\right]  ^{\frac{1}{2\left(  N+1\right)  }} \tag{CI-21l}%
\end{equation}

This expression can be simplified by a huge $N$ using the Stirling's
approximation%
\begin{equation}
\left\vert \alpha_{1}(0)-\alpha_{2}(0)\right\vert \ll\sqrt{2\left(
N+1\right)  } \label{CI-21m}%
\end{equation}
In fact these are the two \textit{macroscopicity} condition: that $\left\vert
\alpha_{1}(0)-\alpha_{2}(0)\right\vert $ and $N$ should be large.
\end{enumerate}

We will consider that $\alpha_{1}(0)-\alpha_{2}(0)$ and $N+1$ always satisfy
these macroscopicity conditions. Then from (\ref{CI-21f}) the basis $\left\{
\left\vert \alpha_{1}(0)\right\rangle ,\left\vert \alpha_{2}(0)\right\rangle
\right\}  $ is quasi-orthogonal and we have
\begin{align}
\rho_{11}^{(ND)}(t)  &  =\left\langle \alpha_{1}(0)\right\vert \rho
^{(ND)}(t)\left\vert \alpha_{1}(0)\right\rangle \nonumber\\
\rho_{12}^{(ND)}(t)  &  =\left\langle \alpha_{1}(0)\right\vert \rho
^{(ND)}(t)\left\vert \alpha_{2}(0)\right\rangle \nonumber\\
\rho_{21}^{(ND)}(t)  &  =\left\langle \alpha_{2}(0)\right\vert \rho
^{(ND)}(t)\left\vert \alpha_{1}(0)\right\rangle \nonumber\\
\rho_{22}^{(ND)}(t)  &  =\left\langle \alpha_{2}(0)\right\vert \rho
^{(ND)}(t)\left\vert \alpha_{2}(0)\right\rangle \label{CI-22}%
\end{align}
then from eq. (\ref{CI-16}) we have%
\begin{align}
\rho_{11}^{(ND)}(t)  &  =ab^{\ast}\left\langle \alpha_{1}(0)|\alpha
_{1}(t)\right\rangle \left\langle \alpha_{2}(t)|\alpha_{1}(0)\right\rangle
+a^{\ast}b\left\langle \alpha_{1}(0)|\alpha_{2}(t)\right\rangle \left\langle
\alpha_{1}(t)|\alpha_{1}(0)\right\rangle \nonumber\\
\rho_{12}^{(ND)}(t)  &  =ab^{\ast}\left\langle \alpha_{1}(0)|\alpha
_{1}(t)\right\rangle \left\langle \alpha_{2}(t)|\alpha_{2}(0)\right\rangle
+a^{\ast}b\left\langle \alpha_{1}(0)|\alpha_{2}(t)\right\rangle \left\langle
\alpha_{1}(t)|\alpha_{2}(0)\right\rangle \nonumber\\
\rho_{21}^{(ND)}(t)  &  =ab^{\ast}\left\langle \alpha_{2}(0)|\alpha
_{1}(t)\right\rangle \left\langle \alpha_{2}(t)|\alpha_{1}(0)\right\rangle
+a^{\ast}b\left\langle \alpha_{2}(0)|\alpha_{2}(t)\right\rangle \left\langle
\alpha_{1}(t)|\alpha_{1}(0)\right\rangle \nonumber\\
\rho_{22}^{(ND)}(t)  &  =ab^{\ast}\left\langle \alpha_{2}(0)|\alpha
_{1}(t)\right\rangle \left\langle \alpha_{2}(t)|\alpha_{2}(0)\right\rangle
+a^{\ast}b\left\langle \alpha_{2}(0)|\alpha_{2}(t)\right\rangle \left\langle
\alpha_{1}(t)|\alpha_{2}(0)\right\rangle \label{CI-23}%
\end{align}
We can compute these products with eq. (\ref{CI-09}).

\begin{description}
\item[-] For $\left\langle \alpha_{1}(0)|\alpha_{1}(t)\right\rangle $ we have
that $\left\vert \psi\right\rangle =\left\vert \alpha_{1}(0)\right\rangle $
and $\left\vert \varphi(t)\right\rangle =\left\vert \alpha_{1}(t)\right\rangle
$, then from (\ref{CI-11}) and since $\alpha_{1}(t)$ is a real number
\[
a_{n}^{\ast}=e^{-\frac{\left\vert \alpha_{1}(0)\right\vert ^{2}}{2}}%
\frac{\left(  \alpha_{1}(0)\right)  ^{n}}{\sqrt{n!}}andb_{n}=e^{-\frac
{\left\vert \alpha_{1}(0)\right\vert ^{2}}{2}}\frac{\left(  \alpha
_{1}(0)\right)  ^{n}}{\sqrt{n!}}%
\]

\item then we have%
\begin{align}
\left\langle \alpha_{1}(0)|\alpha_{1}(t)\right\rangle  &  =e^{-\left\vert
\alpha_{1}(0)\right\vert ^{2}}\sum_{n=0}^{N}\frac{\left(  \left\vert
\alpha_{1}(0)\right\vert ^{2}\right)  ^{n}}{n!}\left(  e^{-i\frac{z_{0}}%
{\hbar}t}\right)  ^{n}\nonumber\\
&  =e^{-\left\vert \alpha_{1}(0)\right\vert ^{2}}e^{\left\vert \alpha
_{1}(0)\right\vert ^{2}e^{-i\frac{z_{0}}{\hbar}t}} \label{CI-25}%
\end{align}

\item[-] For $\left\langle \alpha_{1}(0)|\alpha_{2}(t)\right\rangle $ we have
that $\left\vert \psi\right\rangle =\left\vert \alpha_{1}(0)\right\rangle $
and $\left\vert \varphi(t)\right\rangle =\left\vert \alpha_{2}(t)\right\rangle
$, then from (\ref{CI-11}), (\ref{CI-12}) and since $\alpha_{1}(t)$ and
$\alpha_{2}(t)$ are real numbers
\begin{equation}
a_{n}^{\ast}=e^{-\frac{\left\vert \alpha_{1}(0)\right\vert ^{2}}{2}}%
\frac{\left\vert \alpha_{1}(0)\right\vert ^{n}}{\sqrt{n!}}\text{ and }%
b_{n}=e^{-\frac{\left\vert \alpha_{2}(0)\right\vert ^{2}}{2}}\frac{\left\vert
\alpha_{2}(0)\right\vert ^{n}}{\sqrt{n!}} \label{CI-26}%
\end{equation}
then%
\begin{align}
\left\langle \alpha_{1}(0)|\alpha_{2}(t)\right\rangle  &  =e^{-\frac
{\left\vert \alpha_{1}(0)\right\vert ^{2}+\left\vert \alpha_{2}(0)\right\vert
^{2}}{2}}\sum_{n=0}^{N}\frac{\left(  \left\vert \alpha_{1}(0)\right\vert
\left\vert \alpha_{2}(0)\right\vert \right)  ^{n}}{n!}\left(  e^{-i\frac
{z_{0}}{\hbar}t}\right)  ^{n}\nonumber\\
&  =e^{-\frac{\left\vert \alpha_{1}(0)\right\vert ^{2}+\left\vert \alpha
_{2}(0)\right\vert ^{2}}{2}}e^{\left\vert \alpha_{1}(0)\right\vert \left\vert
\alpha_{2}(0)\right\vert e^{-i\frac{z_{0}}{\hbar}t}} \label{CI-27}%
\end{align}

\item[-] For $\left\langle \alpha_{2}(0)|\alpha_{1}(t)\right\rangle $ we have
that $\left\vert \psi\right\rangle =\left\vert \alpha_{2}(0)\right\rangle $
and $\left\vert \varphi(t)\right\rangle =\left\vert \alpha_{1}(t)\right\rangle
$, then from (\ref{CI-11}), (\ref{CI-12}) and since $\alpha_{1}(t)$ and
$\alpha_{2}(t)$ are real numbers
\begin{equation}
a_{n}^{\ast}=e^{-\frac{\left\vert \alpha_{2}(0)\right\vert ^{2}}{2}}%
\frac{\left(  \alpha_{2}(0)\right)  ^{n}}{\sqrt{n!}}\text{ and }%
b_{n}=e^{-\frac{\left\vert \alpha_{1}(0)\right\vert ^{2}}{2}}\frac{\left(
\alpha_{1}(0)\right)  ^{n}}{\sqrt{n!}} \label{CI-28}%
\end{equation}
then%
\begin{align}
\left\langle \alpha_{2}(0)|\alpha_{1}(t)\right\rangle  &  =e^{-\frac
{\left\vert \alpha_{1}(0)\right\vert ^{2}+\left\vert \alpha_{2}(0)\right\vert
^{2}}{2}}\sum_{n=0}^{N}\frac{\left(  \left\vert \alpha_{1}(0)\right\vert
\left\vert \alpha_{2}(0)\right\vert \right)  ^{n}}{n!}\left(  e^{-i\frac
{z_{0}}{\hbar}t}\right)  ^{n}\nonumber\\
&  =e^{-\frac{\left\vert \alpha_{1}(0)\right\vert ^{2}+\left\vert \alpha
_{2}(0)\right\vert ^{2}}{2}}e^{\left\vert \alpha_{1}(0)\right\vert \left\vert
\alpha_{2}(0)\right\vert e^{-i\frac{z_{0}}{\hbar}t}} \label{CI-29}%
\end{align}

\item[-] For $\left\langle \alpha_{2}(0)|\alpha_{2}(t)\right\rangle $ we have
that $\left\vert \psi\right\rangle =\left\vert \alpha_{2}(0)\right\rangle $
and $\left\vert \varphi(t)\right\rangle =\left\vert \alpha_{2}(t)\right\rangle
$, then from (\ref{CI-12}) and since $\alpha_{2}(t)$ is a real number
\begin{equation}
a_{n}^{\ast}=e^{-\frac{\left\vert \alpha_{2}(0)\right\vert ^{2}}{2}}%
\frac{\left(  \alpha_{2}(0)\right)  ^{n}}{\sqrt{n!}}\text{ and }%
b_{n}=e^{-\frac{\left\vert \alpha_{2}(0)\right\vert ^{2}}{2}}\frac{\left(
\alpha_{2}(0)\right)  ^{n}}{\sqrt{n!}} \label{CI-30}%
\end{equation}
then%
\begin{align}
\left\langle \alpha_{2}(0)|\alpha_{2}(t)\right\rangle  &  =e^{-\left\vert
\alpha_{2}(0)\right\vert ^{2}}\sum_{n=0}^{N}\frac{\left(  \left\vert
\alpha_{2}(0)\right\vert ^{2}\right)  ^{n}}{n!}\left(  e^{-i\frac{z_{0}}%
{\hbar}t}\right)  ^{n}\nonumber\\
&  =e^{-\left\vert \alpha_{2}(0)\right\vert ^{2}}e^{\left\vert \alpha
_{2}(0)\right\vert ^{2}e^{-i\frac{z_{0}}{\hbar}t}} \label{CI-31}%
\end{align}

\end{description}

Now if we consider eqs. (\ref{CI-17}) and (\ref{CI-18}) and remember that the
initial centers of the Gaussians are given by eqs. (\ref{CI-11}) and
(\ref{CI-12}), with no lost of generality we can choose:
\begin{equation}
\alpha_{1}(0)=0 \label{CI-32}%
\end{equation}
and%
\begin{equation}
\alpha_{2}(0)=\frac{m\omega}{\sqrt{2m\hbar^{2}\omega}}L_{0} \label{CI-33}%
\end{equation}

Remember that we have imposed a macroscopic condition to the initial
conditions, i.e. $\left\vert \alpha_{1}(0)-\alpha_{2}(0)\right\vert \gg1$ and
$\left\vert \alpha_{1}(0)-\alpha_{2}(0)\right\vert \ll\left[  2\left(
N+1\right)  !\right]  ^{\frac{1}{2\left(  N+1\right)  }}$. So in the case
given by (\ref{CI-32}) and (\ref{CI-33}) we have%
\begin{equation}
\left\vert \alpha_{1}(0)-\alpha_{2}(0)\right\vert =\alpha_{2}(0)\gg1\text{
\ \ and \ \ }\left\vert \alpha_{1}(0)-\alpha_{2}(0)\right\vert \ll\left[
2\left(  N+1\right)  !\right]  ^{\frac{1}{2\left(  N+1\right)  }}
\label{CI-34}%
\end{equation}
i.e.%
\begin{equation}
\frac{m\omega}{\sqrt{2m\hbar^{2}\omega}}L_{0}\gg1\text{ \ \ and \ }\left[
2\left(  N+1\right)  !\right]  ^{\frac{1}{2\left(  N+1\right)  }}\gg
\text{\ }\frac{m\omega}{\sqrt{2m\hbar^{2}\omega}}L_{0} \label{CI-34b}%
\end{equation}
Then if we substitute (\ref{CI-32}), (\ref{CI-33}) and (\ref{CI-34}) in eq.
(\ref{CI-25}), (\ref{CI-27}), (\ref{CI-29}) and (\ref{CI-31}) and we take into
account (\ref{CI-34})%
\begin{equation}
\left\langle \alpha_{1}(0)|\alpha_{1}(t)\right\rangle =1 \label{CI-35}%
\end{equation}%
\begin{equation}
\left\langle \alpha_{1}(0)|\alpha_{2}(t)\right\rangle =e^{-\frac{\left\vert
\alpha_{2}(0)\right\vert ^{2}}{2}}\cong0 \label{CI-36}%
\end{equation}%
\begin{equation}
\left\langle \alpha_{2}(0)|\alpha_{1}(t)\right\rangle =e^{-\frac{\left\vert
\alpha_{2}(0)\right\vert ^{2}}{2}}\cong0 \label{CI-37}%
\end{equation}%
\begin{equation}
\left\langle \alpha_{2}(0)|\alpha_{2}(t)\right\rangle =e^{-\left\vert
\alpha_{2}(0)\right\vert ^{2}\left(  1-e^{-i\frac{z_{0}}{\hbar}t}\right)  }
\label{CI-38}%
\end{equation}
Then if we substitute (\ref{CI-35}), (\ref{CI-36}), (\ref{CI-37}) and
(\ref{CI-38}) in eq. (\ref{CI-23}) we have%
\begin{align}
\rho_{11}^{(ND)}(t)  &  \cong0\nonumber\\
\rho_{12}^{(ND)}(t)  &  \cong ab^{\ast}e^{-\left\vert \alpha_{2}(0)\right\vert
^{2}\left(  1-e^{-i\frac{z_{0}^{\ast}}{\hbar}t}\right)  }\nonumber\\
\rho_{21}^{(ND)}(t)  &  \cong a^{\ast}be^{-\left\vert \alpha_{2}(0)\right\vert
^{2}\left(  1-e^{-i\frac{z_{0}}{\hbar}t}\right)  }\nonumber\\
\rho_{22}^{(ND)}(t)  &  \cong0 \label{CI-39}%
\end{align}

We see that in the last equation there is an exponential of an exponential,
and if we develop the second exponential and substituting $z_{0}$ for its
value according to eq. (\ref{CI-07}), we have from eq. (\ref{CI-20}) and eq.
(\ref{CI-33})%
\begin{equation}
\rho_{ij}^{(ND)}(t)\propto e^{-\frac{m\omega}{2\hbar^{2}}L_{0}^{2}\gamma_{0}t}
\label{CI-40}%
\end{equation}
So a simple decaying time $t_{R}=\frac{\hbar}{\gamma_{0}}$ is given by the
original pole of eq. (\ref{cuac-03}) but a new decaying pole appears with an
imaginary part%
\begin{equation}
\tilde{\gamma}_{0}=\frac{m\omega}{2\hbar^{2}}L_{0}^{2}\gamma_{0} \label{CI-41}%
\end{equation}
so, the new decaying time is $t_{D}=\frac{\hbar}{\widetilde{\gamma_{0}}}$ or
\begin{equation}
t_{D}=\frac{2\hbar^{2}}{m\omega}\frac{1}{L_{0}^{2}}t_{R} \label{CI-42}%
\end{equation}
the same time was found by Omn\`{e}s in \cite{Omnesazul} or (\ref{131'}) and
corresponds to the definition (\ref{34'}).

In fact, we can recover the same result. In \cite{Omnesazul} the result is
valid for small $t$. In the general case, and considering that $\alpha
_{2}(0)\gg1$, from eqs. (\ref{CI-20}) and (\ref{CI-39})\ we have:%
\begin{align}
\rho^{(ND)}(t)  &  =\left\{  ab^{\ast}\left\vert \alpha_{1}(0)\right\rangle
\left\langle \alpha_{2}(0)\right\vert +ba^{\ast}\left\vert \alpha
_{1}(0)\right\rangle \left\langle \alpha_{2}(0)\right\vert \right\}
\nonumber\\
&  \exp\left[  -\frac{1}{2}\left\vert \alpha_{2}(0)-\alpha_{2}(0)\right\vert
^{2}\left(  1-e^{-\frac{\gamma}{\hbar}t}\right)  \right]  \label{CI-43}%
\end{align}
the same expression that can be found on page 290 of~\cite{Omnesazul}. So the
coincidence of both formalisms is completely proved.

\paragraph{Final remarks.}

i.- As we have said in the macroscopic case the basis \{%
$\vert$%
$\alpha_{1}(0)\rangle,|\alpha_{2}(t)\rangle\}$ is orthogonal and it is the one
defined in section II.F. In fact for $t>t_{D}$ the evolution of $\rho_{R}(t)$
is produced by the p-relevant poles while for $t<t_{D}$ the evolution is
produced by all the poles, either p-relevant and p-irrelevant. Moreover the
corresponding $\rho_{R}(t)$ and $\rho_{P}(t)$ coincide at $t=t_{D},$ with all
their derivatives.

ii.- Let us see what happens if we change the two Gaussian initial condition by%

\begin{equation}
\left\vert \Phi(0)\right\rangle =\left\vert z_{n}\right\rangle \label{CI-46}%
\end{equation}
Then it can be proved that if $\gamma_{m}=n\gamma_{0}<\gamma_{m+1}%
=(n+1)\gamma_{0}$ the decoherence times are $t_{D}=\hbar/\gamma_{n}%
=\hbar/n\gamma_{0}$. Therefore these times change with $n$. These example
shows that the initial condition chooses and the relevant poles define $t_{D}$
as explained in section II.F

Nevertheless in any case we always have that
\begin{equation}
t_{R}=\hbar/\gamma_{0}\text{ and }t_{D}\leq\hbar/\gamma_{1} \label{CI-47}%
\end{equation}

These results coincide with those of Zurek spin model \cite{Paz-Zurek} where
we also have an initial condition dependence. \ This examples shows that there
are many candidates decoherence times and that the initial conditions choose
among them.

\section{\label{Conclusions}Conclusions}

In this paper we have:

i.- Discussed a general scheme of decoherence, that in principle can be used
by many formalisms.

ii.- We have given a quite general definition of a moving preferred basis
$\{\widetilde{|j(t)\rangle}\}$ and of the relaxation time

iii.- We have introduced different characteristic (decaying evolution) times,
and also how the decoherence time is chosen by the initial conditions.

We hope that these general results will produce some light in the general
problem of decoherence.

The Omn\`{e}s formalism, of references \cite{Omnesazul}, \cite{OmnesPh} and
\cite{OmnesRojo}, contains the most general definition of moving preferred
basis of the literature on the subject. Our basis have another conceptual
frame: the catalogue of decaying modes in the non-unitary evolution of a
quantum system. But since the Omn\`{e}s formalism is the best available it is
very important for us to show the coincidence of both formalisms, as we have
done in one model, at least (see section III).

Of course we realize that, to prove our proposal, more examples must be added,
as we will do elsewhere. But we also believe that we have a good point of
depart. In fact, probably the coincidences that we have found in the Omn\`{e}s
model could be a general feature of the decoherence phenomenon. Essentially
because, being the poles catalogue the one that contains \textit{all the
possible decaying modes} of the non unitary evolutions, since relaxation and
decoherence are non-unitary evolutions, necessarily they must be contained
within this catalogue, .

\section{Acknowledgments}

We are very grateful to Roberto Laura, Olimpia Lombardi, Roland Omn\`{e}s and
Maximilian Schlosshauer for many comments and criticisms. This research was
partially supported by grants of the University of Buenos Aires, the CONICET
and the FONCYT of Argentina.

\section{Appendix A.}

\subsection{Observables that see some poles.}

In this appendix we will introduce a particular example of observables, of the
same system, such that some observables would see some poles while other would
see other ones. Essentially it is a bi-Friedrich-model.

Let us consider a system $\mathcal{S}$ with Hamiltonian:%
\[
H=H_{0}+H_{Int}%
\]
where%
\[
H_{0}=\Omega_{1}|1\rangle\langle1|+\Omega_{2}|2\rangle\langle2|+2\int
_{0}^{\infty}\omega|\omega\rangle\langle\omega|d\omega
\]
and
\[
H_{Int}=\int_{0}^{a}V_{\omega}^{(1)}\left[  |\omega\rangle\langle
1|+|1\rangle\langle\omega|\right]  d\omega+\int_{b}^{\infty}V_{\omega^{\prime
}}^{(2)}\left[  |\omega^{\prime}\rangle\langle2|+|2\rangle\langle
\omega^{\prime}|\right]  d\omega
\]
where $a<b$ and $\langle1|2\rangle=\langle\omega|2\rangle=\langle
1|\omega\rangle=0.$ This Hamiltonian can also reads:%
\[
H=H_{1}+H_{2}%
\]
where%
\[
H_{1}=\Omega_{1}|1\rangle\langle1|+\int_{0}^{\infty}\omega|\omega
\rangle\langle\omega|d\omega+\int_{0}^{a}V_{\omega}^{(1)}\left[
|\omega\rangle\langle2|+|2\rangle\langle\omega|\right]  d\omega
\]
and%
\[
H_{2}=\Omega_{2}|2\rangle\langle2|+\int_{0}^{\infty}\omega^{\prime}%
|\omega^{\prime}\rangle\langle\omega^{\prime}|d\omega^{\prime}+\int
_{b}^{\infty}V_{\omega^{\prime}}^{(2)}\left[  |\omega^{\prime}\rangle
\langle1|+|1\rangle\langle\omega^{\prime}|\right]  d\omega^{\prime}%
\]
Then it is easy to prove that%
\[
\lbrack H_{1},H_{2}]=0
\]
and that%
\[
\exp(-\frac{i}{\hbar}Ht)=\exp(-\frac{i}{\hbar}H_{1}t)\exp(-\frac{i}{\hbar
}H_{2}t)
\]
Let us now decompose the system as $\mathcal{S=P}_{1}\cup\mathcal{P}_{2}$
\ where part $\mathcal{P}_{1}$ is related with Hamiltonian $H_{1}$ and part
$\mathcal{P}_{2}$ related with Hamiltonian $H_{2}$. Let us observe that these
two parts are not independent since they share a common continuous spectrum,
i. e. $2\int_{0}^{\infty}\omega|\omega\rangle\langle\omega|d\omega$. Moreover
let the corresponding relevant observable spaces be $\mathcal{O}_{1}\otimes
I_{E1}$ for $\mathcal{P}_{1}$ and $\mathcal{O}_{2}\otimes I_{E2}$ for
$\mathcal{P}_{2},$ where $\mathcal{O}_{1}$ has basis \{$|1\rangle\},$ and
$\mathcal{O}_{E1}$ has basis \{$|\omega\rangle\}$ while $\mathcal{O}_{2}$ has
basis \{$|2\rangle\},$ and $\mathcal{O}_{E2}$ basis \{$|\omega^{\prime}%
\rangle\}.$ Moreover let us consider the two relevant observables of system
$\mathcal{S=P}_{1}\cup\mathcal{P}_{2}$
\[
\mathbb{O}_{i}=O_{1}\otimes I_{E1}\otimes\mathcal{I}_{2}\otimes I_{E2}\text{
and }\mathbb{O}_{2}=I_{1}\otimes I_{E1}\otimes O_{2}\otimes I_{E2}%
\]
where the $I$ are the corresponding unit operators. Then
\[
(\rho(t)|\mathbb{O}_{1})=(\rho(0)|\exp(\frac{i}{\hbar}H_{1}t)O_{1}\exp
(-\frac{i}{\hbar}H_{1}t)\otimes I_{E1}\otimes\mathcal{I}_{2}\otimes I_{E2})
\]%
\[
(\rho(t)|\mathbb{O}_{2})=(\rho(0)|I_{1}\otimes I_{E1}\otimes\exp(\frac
{i}{\hbar}H_{1}t)O_{2}\exp(.\frac{i}{\hbar}H_{2}t)\otimes I_{E2})
\]
and therefore $\mathbb{O}_{1}$ only sees the evolution in part $\mathcal{P}%
_{1}$ while $\mathbb{O}_{2}$ only sees the evolution in part $\mathcal{P}%
_{2}.$ Then, since the poles of part $\mathcal{P}_{1}$ correspond to the
decaying modes of the evolution of this part (and we know that the Friedrich
model of this subsystem generically do have poles) $\mathbb{O}_{1}$ only sees
the poles of part $\mathcal{P}_{1}$. Respectively $\mathbb{O}_{2}$ only sees
the poles of part $\mathcal{P}_{2}.$ q. e. d.

Now we can consider that the poles of part $\mathcal{P}_{1}$ define a
relaxation time $t_{R1\text{ }}$while the poles of part $\mathcal{P}_{2}$
define a relaxation time $t_{R2\text{ }}.$ If $t_{R1\text{ }}\ll t_{R2\text{
}}$part $\mathcal{P}_{1}$ decoheres and becomes classical in a\ short time
$t>t_{R1\text{ }}$while part $\mathcal{P}_{2}$ remains quantum for a large
time $t_{\text{ }}<t_{R2\text{ }}.$ Then for $t$ such that $t_{R1\text{ }%
}<t<t_{R2\text{ }}$ part $\mathcal{P}_{1}$ behaves

classically while part $\mathcal{P}_{2}$ remains quantum. Precisely: system
$\mathcal{S}$ observed by $\mathbb{O}_{i}=O_{1}\otimes I_{E1}\otimes
\mathcal{I}_{2}\otimes I_{E2}$ seems classical while observed by
$\mathbb{O}_{2}=I_{1}\otimes I_{E1}\otimes O_{2}\otimes I_{E2}$ \ seems
quantum. In fact this is the behavior of a generic physical system.


\begin{thebibliography}{99}                                                                                               %


\bibitem {CQG-CFLL-08}M. Castagnino, S. Fortin, R. Laura and O. Lombardi,
\textit{Class. Quantum Grav.}, \textbf{25}, 154002, 2008.

\bibitem {Omnes}R. Omn\'{e}s, \textit{Brazilian Journal of Physics,
}\textbf{35, }207, 2005.

\bibitem {Bub}J. Bub, \textit{Interpreting the Quantum World}, Cambridge
University Press, Cambridge, 1997.

\bibitem {Max}M. Schlosshauer, \textit{Decoherence and the
Quantum-to-Classical transition}, Springer, Berlin, 2007.

\bibitem {GP}R. Gambini, J. Pulin, \textit{Found. of Phys.}, \textbf{37}, 7, 2007.

R. Gambini, R. A. Porto and J. Pulin, \textit{Gen. Rel. Grav.}, \textbf{39},
8, 2007.

R. Gambini and J. Pulin, \textquotedblleft Modern space-time and
undecidability\textquotedblright, in V. Petkov (ed.), \textit{Fundamental
Theories of Physics (Minkowski Spacetime: A Hundred Years Later)}, Vol. 165,
Springer, Heidelberg, 2010.

\bibitem {connel}G. Ford, R. O'Connel, \textit{Phys. Rev. Lett. A},
\textbf{286},87,2001.

\bibitem {CP}G. Casati, T. Prosen, \textit{Phys. Rev. A}, \textbf{72}, 032111, 2005.

\bibitem {SID}M. Castagnino\textit{,} \textit{Int. J. Theor. Phys.,}
\textbf{38}, 1333, 1999.

M. Castagnino and R. Laura, \textit{Phys. Rev. A}, \textbf{62}, 022107, 2000.

\bibitem {SID'}M. Castagnino and A. Ordo\~{n}ez, \textit{Int. J. Theor.
Phys.}, \textbf{43}, 695, 2004.

G. Murgida, M. Castagnino, \textit{Physica A}, \textbf{381}, 170, 2007.

\bibitem {DT}M. Castagnino and O. Lombardi, \textit{Phys. Rev. A},
\textbf{72}, 012102, 2005.

\bibitem {MHI}O. Lombardi, M. Castagnino. \textit{Studies in History and
Philosophy of Modern Physics}, \textbf{39}, 380, 2008.

\bibitem {PLA}M. Castagnino, \textit{Physics Letters A}, \textbf{357}, 97, 2006.

\bibitem {JPA2}M. Castagnino, S. Fortin, and O. Lombardi,\textit{\ J. Phys. A:
Math. Theor.}, \textbf{43}, 065304, 2010.

\bibitem {Studies}M. Castagnino, O. Lombardi\textit{, } \textit{Studies in
History and Philosophy of Modern Physics}, \textbf{35}, 73, 2004.

\bibitem {MPLA}M. Castagnino, S. Fortin and O. Lombardi, \textit{Mod. Phys.
Lett. A}, \textbf{25, }1431, 2010.

\bibitem {OmnesPh}R. Omn\`{e}s, \textit{Phys. Rev}. \textit{A}, \textbf{56},
3383-3394, 1997, \textbf{65}, 052119, 2002.

\bibitem {OmnesRojo}R. Omn\`{e}s, \textit{Understanding quantum mechanics,
}Princeton Univ. Press, Princeton, 1999.

\bibitem {JPA}M. Castagnino, R. Id Betan, R. Laura, R. J. Liotta, \textit{J.
Phys. A: Math. Gen.}, \textbf{35}, 6055-6074, 2002.

\bibitem {Bohm}A. Bohm, \textit{Quantum mechanics, foundations and
applications,}Springer Verlag, Berlin, 1986.

\bibitem {Weder}R. Weder, \textit{Jour. Math. Phys., }\textbf{15,} 20, 1974.

\bibitem {Sudarsham}E. Sudarsham, C. Chiu, V. Gorini, \textit{Phys. Rev. D},
\textbf{18,} 2914, 1978.

\bibitem {PRA1}M. Castagnino and R. Laura, \textit{Phys. Rev. A}, \textbf{56,}
108, 1997.

\bibitem {PRA2}R. Laura and M. Castagnino\textit{,} \textit{Phys. Rev. A},
\textbf{57}, 4140, 1998.

\bibitem {PRE}R. Laura and M. Castagnino, \textit{Phys. Rev. E}, \textbf{57},
3948, 1998.

\bibitem {Khalfin}K. Urbanowski, \textit{Eur. Phys. J. D}, \textbf{54}, 25, 2009

K. Urbanowski, J. Piskorski, arXiv:0908.2219v2, 2009.

\bibitem {BH}N. Bleistein, R. Handelsman, \textit{Asymptotic expansion of
integrals, }Dover Inc., New York, 1986.

\bibitem {KhalfinEx}C. Rothe, S. I. Hintschich, and A. P. Monkman,
\textit{Phys. Rev. Lett.}, \textbf{96}, 163601, 2006.

\bibitem {Cos}M. \ Castagnino, F. Gailoli, E. Gunzig, \textit{Foud. Cosmic
Phys., }\textbf{16}\textit{, }221-375, 1996.

\bibitem {K1}F. Gaioli, E. Garc\'{\i}a-\'{A}lvarez, J. Guevara, \textit{Int.
J. Theor. Phys., }\textbf{36,} 2167-2219, 1997.

\bibitem {K2}D. Arb\'{o}, M. Castagnino, F. Gaioli, S. Iguri, \textit{Physica
A, }\textbf{274, }469-495\textbf{.}

\bibitem {Manoloetal}F. Facchi, M. Gadella, S. Pascazio, G. Pronko
\textquotedblleft The Friedrich Model and its use in resonance
phenomena\textquotedblright, private communication, 2008.

\bibitem {Omnesazul}R. Omn\`{e}s, \textit{The Interpretation of Quantum
Mechanics, }Princeton University Press, Princeton, 1994.

\bibitem {Paz-Zurek}P. Paz and W. H. Zurek, \textquotedblleft
Environment-induced decoherence and the transition from quantum to
classical\textquotedblright\textit{, }in D. Heiss (ed.), \textit{Lecture Notes
in Physics, Vol. 587}, Springer, Heidelberg, 2002.

\bibitem {Paz-Habib-Zurek}J. P. Paz, S. Habib and W. H. Zurek,
\textit{Physical Review D}, \textbf{47}, 488, 1993.

\bibitem {Ex}E. Joos, H. D. Zeh, C. Kiefer, D. Giulini, J. Kupsch and I. O.
Stamatescu, \textit{Decoherence and the Appearance of a Classical World in
Quantum Theory, }Springer Verlag, Berlin, 2003.

\bibitem {LCIB}R. Laura, M. Castagnino, R. Id Betan, \textit{Physica A,
}\textbf{271, }357-386, 1999.
\end{thebibliography}
\end{document}